\documentclass[letterpaper,twocolumn,10pt]{article}
\usepackage{usenix-2020-09}

\usepackage{tikz}       
\usetikzlibrary{arrows.meta}
\usepackage{amsmath}

\usepackage{graphicx}
\usepackage{booktabs}
\usepackage{algorithm}
\usepackage{algpseudocode}
\usepackage{array}
\usepackage{tabularx}
\usepackage{xspace}
\usepackage{subcaption}
\usepackage{pifont}   
\usepackage{enumitem}
\newlist{tightitem}{itemize}{1}
\setlist[tightitem]{label=\textbullet,leftmargin=1.2em,labelsep=0.4em,topsep=2pt,itemsep=2pt,parsep=0pt,partopsep=0pt}
\newlist{tightenum}{enumerate}{1}
\setlist[tightenum]{label=\arabic*.,leftmargin=1.5em,labelsep=0.4em,topsep=2pt,itemsep=2pt,parsep=0pt,partopsep=0pt}

\newcommand{\tyes}{\ding{51}}

\newcommand{\ignore}[1]{}

\newcommand{\pgheading}[1]{\noindent\textbf{#1.}}

\usepackage{titlesec}
\usepackage{enumitem}
\usepackage[font=small]{caption}

\setlist{
  nosep,
  topsep=2pt,
  partopsep=0pt,
  parsep=0pt,
  itemsep=1pt
}

\newcommand{\tracea}{FreeInference Trace\xspace}   
\newcommand{\traceb}{Chutes Trace\xspace}   

\newcommand{\sys}{PrefixSim\xspace}

\newcommand{\model}{Qwen3-Coder-30B\xspace}

\newcommand{\findingbox}[1]{%
\vspace{0.5em}
\noindent\fbox{%
    \parbox{0.97\linewidth}{%
        \textbf{Finding \& Implication:} #1
    }%
}
\vspace{0.5em}
}

\begin{document}

\date{}
\title{\Large \bf When Fancy Eviction Fails: Rethinking Cache Replacement For LLM Prefix Reuse}

\author{
{\rm Yiyu Liu}\\
Harvard University
\and
{\rm Minlan Yu}\\
Harvard University
\and
{\rm Juncheng Yang}\\
Harvard University
}

\maketitle

\begin{abstract}
Long-running LLM applications and autonomous agents repeatedly send growing context, making prefix caching critical for reducing prefill cost. This raises a fundamental question: do decades of traditional cache-management techniques transfer directly to agentic and conversational LLM serving? We analyze production traces comprising over 20 billion processed tokens and evaluate 14 eviction algorithms across capacity-constrained HBM and large memory-pool settings. We find that simple Least Recently Used (LRU) is surprisingly robust, offering competitive performance against sophisticated traditional policies, though substantial headroom to Belady's optimum remains. The reason is structural: prefix reuse is dominated by the steady pacing of active sessions, making recency an unusually strong predictor for eviction. Nevertheless, prefix caching diverges from traditional workloads in ways that recency ignores, notably the prevalence of single-turn requests, extreme session footprint skew, and highly non-uniform miss costs because attention computation scales with token depth. To close the gap to optimality, we establish several design principles for prefix-cache management: keeping recency as the baseline while incorporating quick demotion to filter one-hit prompts, compute-aware eviction to prioritize expensive misses, and capacity-adaptive eviction granularity. Our traces and simulator are publicly available.
\end{abstract}

\section{Introduction}
%
\definecolor{ovfind}{HTML}{4C78A8}
\definecolor{ovimp}{HTML}{A8802E}
\definecolor{ovdesign}{HTML}{54A24B}

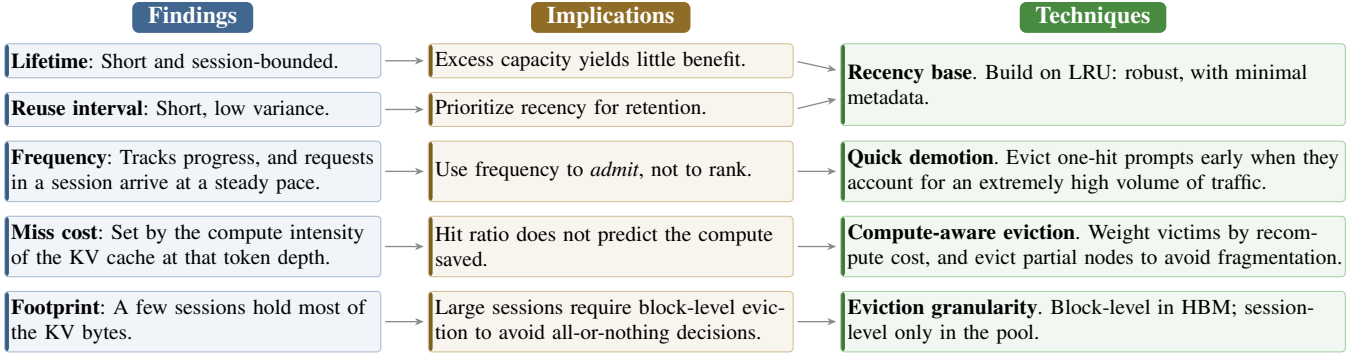
\begin{figure*}[t]
  \centering
  \begin{tikzpicture}[
      font=\footnotesize,
      box/.style={rounded corners=1.5pt, line width=0.4pt, align=left,
                  minimum height=0.80cm, inner sep=2.5pt, anchor=north west},
      fbox/.style={box, draw=ovfind!50, fill=ovfind!8, text width=4.80cm,
                   path picture={\fill[ovfind!75!black]
                    (path picture bounding box.north west) rectangle
                    ([xshift=1.6pt]path picture bounding box.south west);}},
      ibox/.style={box, draw=ovimp!50, fill=ovimp!8, text width=4.65cm,
                   path picture={\fill[ovimp!75!black]
                    (path picture bounding box.north west) rectangle
                    ([xshift=1.6pt]path picture bounding box.south west);}},
      dbox/.style={box, draw=ovdesign!50, fill=ovdesign!8, text width=6.5cm,
                   path picture={\fill[ovdesign!75!black]
                    (path picture bounding box.north west) rectangle
                    ([xshift=1.6pt]path picture bounding box.south west);}},
      chip/.style={font=\small\bfseries, text=white, anchor=south,
                   rounded corners=2.5pt, inner xsep=6pt, inner ysep=1.8pt},
      arr/.style={-{Stealth[length=3.4pt,width=2.8pt]}, draw=black!45,
                  line width=0.5pt, shorten >=1.5pt, shorten <=1.5pt},
    ]
    \def\cf{0}    \def\ci{5.605} \def\cd{11.06}
    \coordinate (cI) at (\ci,0);
    \coordinate (cD) at (\cd,0);
    \def\g{0.18cm}

    \node[fbox, minimum height=0.45cm] (F1) at (\cf,0) {\textbf{Lifetime}: Short and session-bounded.};
    \node[fbox, minimum height=0.45cm] (F2) at ([yshift=-\g]F1.south west) {\textbf{Reuse interval}: Short, low variance.};
    \node[fbox] (F3) at ([yshift=-\g]F2.south west) {\textbf{Frequency}: Tracks progress, and requests in a
      session arrive at a steady pace.};
    \node[fbox] (F4) at ([yshift=-\g]F3.south west) {\textbf{Miss cost}: Set by the compute intensity of the
      KV cache at that token depth.};
    \node[fbox] (F5) at ([yshift=-\g]F4.south west) {\textbf{Footprint}: A few sessions hold most of the KV bytes.};

    \node[ibox, minimum height=0.45cm] (I1) at (F1.north -| cI) {Excess capacity yields little benefit.};
    \node[ibox, minimum height=0.45cm] (I2) at (F2.north -| cI) {Prioritize recency for retention.};
    \node[ibox] (I3) at (F3.north -| cI) {Use frequency to \emph{admit}, not to rank.};
    \node[ibox] (I4) at (F4.north -| cI) {Hit ratio does not predict the compute saved.};
    \node[ibox] (I5) at (F5.north -| cI) {Large sessions require block-level eviction to avoid all-or-nothing decisions.};

    \node[dbox, minimum height=1.08cm] (D0) at (F1.north -| cD)
      {\textbf{Recency base}. Build on LRU: robust, with minimal metadata.};
    \node[dbox] (D1) at (F3.north -| cD) {\textbf{Quick demotion}. Evict one-hit prompts early when they
      account for an extremely high volume of traffic.};
    \node[dbox] (D2) at (F4.north -| cD) {\textbf{Compute-aware eviction}. Weight victims by recompute cost, and
      evict partial nodes to avoid fragmentation.};
    \node[dbox] (D3) at (F5.north -| cD) {\textbf{Eviction granularity}. Block-level in HBM; session-level
      only in the pool.};

    \node[chip, fill=ovfind!85!black]   (HF) at ([yshift=0.13cm]F1.north) {Findings};
    \node[chip, fill=ovimp!85!black]    (HI) at ([yshift=0.13cm]I1.north) {Implications};
    \node[chip, fill=ovdesign!85!black] (HD) at ([yshift=0.13cm]D0.north) {Techniques};

    \foreach \i in {1,2,3,4,5} { \draw[arr] (F\i.east) -- (I\i.west); }

    \draw[arr] (I1.east) -- ([yshift=0.18cm]D0.west);
    \draw[arr] (I2.east) -- ([yshift=-0.18cm]D0.west);
    \draw[arr] (I3.east) -- (D1.west);
    \draw[arr] (I4.east) -- (D2.west);
    \draw[arr] (I5.east) -- (D3.west);
  \end{tikzpicture}
  \captionsetup{skip=8pt}
  \caption{Paper overview: We characterize five fundamental properties of production prefix-cache workloads (\S\ref{sec:analysis}), derive their implications for cache management, and map them to our targeted design techniques (\S\ref{sec:designs}).}
  \label{fig:overview}
  \vspace{-0.8em}
\end{figure*}

Long-running conversations and autonomous agents are rapidly pushing LLM context lengths into the hundreds of thousands of tokens~\cite{yao2023react,park2023generative,purav2025manus,qwen3technicalreport,wang2025kvcache}. Recomputing these contexts is increasingly expensive because Transformer prefill cost grows rapidly with sequence length~\cite{vaswani2017attention,kaplan2020scaling}. Fortunately, successive requests within a session typically reuse a long prefix~\cite{gao2024cachedattention}. Modern serving systems therefore rely on \emph{prefix caching}: they retain the key-value (KV) states of previously processed tokens and reuse them across requests, avoiding redundant prefill computation~\cite{kwon2023efficient,zheng2024sglang,qin2025mooncake}. As contexts grow, the efficiency of this cache becomes an increasingly important determinant of LLM serving cost and performance. 

Traditional cache workloads, including web content and block storage, have been studied extensively~\cite{breslau1999web,berger2017adaptsize,kavalanekar2008characterization,li2020cloudblock}, yielding a rich set of eviction algorithms~\cite{cherkasova1998improving,megiddo2003arc,eisenman2019flashield,berg2020cachelib,mcallister2021kangaroo,zhang2024sieve}. Revisiting these algorithms leads to a natural question: \emph{what makes a good eviction policy for LLM prefix caches?} Recent work has begun to characterize aggregate LLM traffic~\cite{liu2026agentic,zhu2026tracelab,wang2025kvcache,aubakirova2026state}, but the access patterns that determine prefix-cache eviction, particularly for agentic workloads, remain largely unexplored. To address this gap, we collect production traces comprising over 20 billion processed tokens from two distinct organizations, and build a high-fidelity simulator to evaluate 14 advanced eviction algorithms spanning recency, frequency, and learned techniques.


We first test whether decades of cache-management techniques transfer directly. Most LLM serving systems use Least Recently Used (LRU) eviction~\cite{kwon2023efficient,zheng2024sglang,qin2025mooncake,liu2025lmcache}, both in capacity-constrained \emph{per-replica HBM} (24--120\,GiB per GPU) and in much larger \emph{global memory pools} (0.25--12\,TiB)~\cite{qin2025mooncake,liu2025lmcache,xie2025strata}. In traditional caches, algorithms that combine recency, frequency, and learned signals can substantially outperform LRU~\cite{megiddo2003arc,jiang2002lirs,zhang2024sieve,yang2023fifo,einziger2017tinylfu,beckmann2018lhd,vietri2018lecar,song2020learning,zhou20253lcache}. We evaluate 14 advanced algorithms~\cite{wang2025kvcache,shi2026asymcache} across both memory regimes and find a surprising result: most offer little improvement over LRU, while frequency-based algorithms often perform substantially worse. At the same time, Belady's optimum~\cite{belady1966study} shows substantial remaining headroom.

This gap suggests that prefix-cache workloads differ fundamentally from the workloads these algorithms were designed for. Understanding why LRU is so difficult to beat---and what signals remain exploitable beyond recency---is therefore the key to designing better prefix-cache management.

Comparing prefix-cache traces with the web and block-storage workloads that drove conventional eviction policies reveals the source of this mismatch. Most prefix reuse comes from successive turns within the same session. As a result, cached blocks have short, session-bounded lifetimes, and reuse intervals are tightly concentrated. Recency captures this structure naturally. Frequency, in contrast, largely reflects how long a session has been active and provides little information about when its blocks will be reused. This explains both LRU's robustness and the poor performance of frequency-based policies.

Beyond these temporal access patterns, prefix caching diverges from traditional caching in two structural ways. First, miss costs are highly non-uniform~\cite{shi2026asymcache}: because attention cost increases with token position, reconstructing a block deep in a prompt is more expensive than reconstructing an equally sized block near the beginning. Thus, two policies with the same hit ratio can incur different prefill costs and time-to-first-token (TTFT). Second, session KV footprints are highly skewed, with a small number of large sessions consuming a disproportionate fraction of cache capacity and contributing to a large portion of hits (i.e., size correlated with frequency). Figure~\ref{fig:overview} summarizes these properties and their implications for cache design.

These observations suggest a simple principle: keep recency as the foundational baseline, and introduce additional signals only to target specific workload characteristics that recency ignores, such as the prevalence of single-turn requests, varying computation costs, or extreme footprint skew.
For example, we find that \emph{quick demotion}~\cite{yang2023fifo} helps primarily when one-hit prompts are common. For non-uniform miss costs, we derive an offline optimum and an efficient approximation, showing that compute-aware eviction can improve performance but must control fragmentation. Finally, the right eviction granularity depends on cache capacity: constrained HBM benefits from block-level management, whereas large memory pools can use session-level eviction to reduce metadata overhead with little efficiency loss.

This paper makes the following main contributions:

\begin{tightitem}

\item \textbf{The first production study of prefix-cache eviction.} We introduce two production traces and \sys, a faithful C++ simulator, to systematically evaluate eviction policies across HBM and memory-pool regimes. We validate our findings on four additional Qwen Bailian workloads~\cite{wang2025kvcache}. Both traces and the simulator are publicly available.\footnote{Simulator: \url{https://github.com/cacheMon/PrefixSim}. Traces: \url{https://huggingface.co/datasets/harvardMadsys/2026_prefixcache}.}

\item \textbf{Structural insights into prefix-cache access patterns.} By contrasting prefix-cache traces with canonical web and block-storage workloads, we identify the structural properties that make recency unusually predictive and render many traditional eviction signals ineffective.

\item \textbf{New design principles for prefix-cache management.} We introduce the \emph{compute-savings ratio} to capture the highly non-uniform cost of prefix-cache misses, and derive an exact offline optimum and an efficient approximation that bound achievable performance. Guided by both our structural findings and these analytical bounds, we show when and how to augment recency with quick demotion, compute-aware eviction, and capacity-adaptive eviction granularity, yielding concrete guidance for both constrained HBM caches and large global memory pools.
\end{tightitem}

\section{Background}

\subsection{LLM inferences and KV Cache}

During LLM inference, Transformer models~\cite{vaswani2017attention} generate output sequentially, using an attention mechanism that requires computing relationships against all preceding tokens. To eliminate the redundant computation of re-evaluating the entire sequence at every step, systems rely on a \textit{KV cache}~\cite{kwon2023efficient} to store previously computed key and value states. This cache divides the inference process into two distinct phases: \textit{prefill} and \textit{decode}.
During the prefill phase, the model processes the entire prompt,
saving the resulting key and value vectors to GPU memory.
The latency of this initial processing, including any queuing delay,
is measured as Time To First Token (TTFT).
In the subsequent decode phase, the model generates tokens sequentially,
attending to the KV cache built during the prefill phase rather than recomputing them.
The speed of this generation is measured as the Time Per Output Token (TPOT).
Because the KV cache must preserve the states of all active tokens,
it rapidly becomes a severe memory bottleneck as context lengths and batch sizes grow,
fundamentally constraining serving capacity~\cite{kwon2023efficient}.

\subsection{LLM Serving Prefix Cache}
\label{sec:background:prefix-caching}

In many real-world workloads, requests frequently share common prefixes,
such as lengthy system prompts~\cite{park2023generative,purav2025manus},
few-shot examples~\cite{brown2020language}, and, in multi-turn conversations and agent
loops, the whole history of the exchange so far~\cite{yao2023react}.

Recomputing these shared prefixes for every request is highly inefficient
because the computational cost of the prefill phase grows rapidly.
As context lengths expand to support document question
answering~\cite{bai2024longbench,liu2024lost} and autonomous
agents~\cite{yao2023react,park2023generative}, the compute burden becomes massive.
Moreover, for the full exact self-attention layers,
the computation scales quadratically with sequence length~\cite{vaswani2017attention,kaplan2020scaling}.
The rise of Mixture-of-Experts (MoE)
models~\cite{zhou2022mixture,shen2023mixture,qwen3technicalreport} exacerbates
this quadratic bottleneck by sparsifying MLP computations,
causing attention to dominate the prefill phase and making cache hits even more valuable.

To eliminate massive redundant computation, serving systems cache previously computed KV states in prefix or radix trees~\cite{zheng2024sglang}.
Because a token's KV state depends on all preceding tokens, a cached state can be safely reused if and only if its entire preceding sequence exactly matches the new request.

Recognizing these substantial benefits, modern LLM serving systems have widely
adopted this \emph{prefix caching} technique.
In fact, production deployments report that a large portion of their total traffic
is now served directly from these caches~\cite{qin2025mooncake,wang2025kvcache}.
These systems efficiently manage KV states by chunking them into fixed-length blocks
(typically 16 tokens per block~\cite{kwon2023efficient}).
To maximize hit rates, they typically deploy prefix caches across a two-tiered memory hierarchy.
At the first tier, the serving engine keeps KV blocks in the
GPU High-Bandwidth Memory (HBM)~\cite{zheng2024sglang,kwon2023efficient}.
At the second tier, systems like
Mooncake~\cite{qin2025mooncake} and Strata~\cite{xie2025strata} introduce a massive
global pool backed by host DRAM and SSDs. This shared pool is one to two orders of
magnitude larger than the local HBM, trading the overhead of cross-device data transfers
for vastly increased capacity.

While the prefix tree provides an effective logical data structure for organizing and matching
these shared KV blocks, it does not dictate physical cache eviction.
As long as the system can determine that a cached block shares exactly the same prefix as
the new request, that block can be safely reused.
Hence, physical cache eviction is fundamentally decoupled from the logical prefix tree
structure, allowing algorithms to drop blocks from anywhere in the cache.

\subsection{Sessions}
\label{sec:background:session}

A \emph{session} represents a single continuous conversation~\cite{characterai2024optimizing} or autonomous agent task~\cite{yao2023react,park2023generative}.
It comprises sequential requests (\emph{turns}) that progressively append to a shared context.
While engines schedule individual requests, the session remains the logical unit of interaction.

This append-only structure preserves history and maximizes prefix cache reuse.
Each request appends new information, such as user inputs or tool outputs, to an earlier turn.
This iterative loop, where a model dispatches a tool call, the runtime executes it, and the output is seamlessly woven back into the context for the next turn, forms the agent workflow.

\subsection{Traditional Web and Storage Cache}
\label{sec:background:traditional}
Traditional caching research evaluates policies against a broad spectrum of workloads to capture diverse access patterns. Major representative classes include in-memory key-value caches (e.g., Twitter Twemcache~\cite{yang2020twemcache} and Meta Memcached~\cite{berg2020cachelib}), flash caches~\cite{eisenman2019flashield,mcallister2021kangaroo}, content delivery networks (CDNs)~\cite{berger2017adaptsize}, and cloud block storage~\cite{li2020cloudblock,kavalanekar2008characterization}. In these environments, object \emph{lifetimes} and \emph{reuse intervals} are typically driven by long-term global popularity spanning days or weeks. Popular objects experience high request concurrency, making historical frequency a reliable predictor of future reuse~\cite{breslau1999web,einziger2017tinylfu}. Furthermore, \emph{miss costs} are generally uniform or only related to object size~\cite{berger2017adaptsize,beckmann2018lhd}. Decades of eviction algorithms have been heavily optimized for these specific behaviors~\cite{megiddo2003arc,jiang2002lirs,einziger2017tinylfu,beckmann2018lhd,zhang2024sieve}, including recent state-of-the-art policies like S3-FIFO~\cite{yang2023fifo} and SIEVE~\cite{zhang2024sieve}.

To contextualize how LLM serving departs from these traditional assumptions, our analysis in \S\ref{sec:analysis} (Table~\ref{tab:workloads}) contrasts our LLM prefix traces with two representative traditional workloads from cacheMon~\cite{cachemon}: \emph{Wikipedia}, a CDN request log for article text encompassing 207.6\,M requests over 18.4\,M objects across 21.0 days; and \emph{CloudPhysics w01}~\cite{waldspurger2015shards}, representing disk I/O of a production VM volume with 215.8\,M requests over 122.8\,M objects across 6.3 days.


\section{Fancy Algorithms Do Not Work Well}
\label{sec:algorithms}

\begin{table}[t]
  \centering
  \caption{LLM inference request traces from two organizations. Requests are capped at 256K tokens. }
  \label{tab:traces}
  \vspace{-0.8em}
  \small
  \begin{tabular}{l rr}
    \toprule
     & \tracea & \traceb \\
    \midrule
    Requests                     & 327.5\,K & 515.8\,K \\
    \quad in multi-turn sessions &  34.3\%  &  26.1\%  \\
    Span                         &   7.0\,d & 130.9\,d \\
    \midrule
    Tokens processed             &  10.5\,B &   9.7\,B \\
    Unique tokens                &  0.63\,B &  2.23\,B \\
    \midrule
    Request tokens, average      &  32.0\,K &  18.7\,K \\
    \quad 99th percentile        & 219.3\,K &  98.3\,K \\
    \bottomrule
  \end{tabular}
\end{table}

\begin{table}[t]
  \centering
  \caption{Cache sizes we evaluate.}
  \label{tab:cache-points}
  \vspace{-0.8em}
  \small
  \begin{tabular}{r r >{\raggedright\arraybackslash}p{4.1cm}}
    \toprule
    Size & Blocks & Example setting \\
    \midrule
    \multicolumn{3}{@{}l}{\emph{HBM, per replica: 24--120\,GiB}} \\
    {24\,GiB} & 16{,}384 & native-context floor \\
    {48\,GiB} & 32{,}768 & 1$\times$H200 at 0.8 memory fraction \\
    {96\,GiB} & 65{,}536 & 1$\times$B200, bf16 weights \\
    \midrule
    \multicolumn{3}{@{}l}{\emph{DRAM/SSD pool, global: 0.25--12\,TiB}} \\
    {1\,TiB} & 699{,}051 & full node in Mooncake~\cite{qin2025mooncake} \\
    \bottomrule
  \end{tabular}
  \\\vspace{0.5em}
\end{table}

\begin{table*}[t]
  \centering
  \caption{Comparison of public LLM-serving traces with the two traces introduced in this work. Only \tracea
  and \traceb combine fine-grained block identity scoped across
  sessions, agentic traffic, and a multi-day span.}
  \label{tab:trace-compare}
  \vspace{-0.5em}
  \small
  \begin{tabular}{@{}l r r l l l c@{}}
    \toprule
    Trace & Requests & Tokens & Span & Block size & Scope & Agentic \\
    \midrule
    TraceLab~\cite{zhu2026tracelab}$^{\dagger}$ & 357\,K & 54.9\,B & 253\,d, scattered & none & --- & \tyes \\
    Mooncake, tool \& agent~\cite{qin2025mooncake} & 23.6\,K & 0.20\,B & 1\,h & coarse (512\,tokens) & global & \tyes \\
    AgentX~\cite{semianalysis2026agentx} & 98.8\,K & 21.6\,B & per session & intermediate (64\,tokens) & per session & \tyes \\
    Qwen Bailian~\cite{wang2025kvcache}$^{*}$ & 10.8--172.8\,K & 0.05--0.25\,B & 2\,h & fine (16\,tokens) & global & very little \\
    \midrule
    \tracea (this work) & 327.5\,K & 10.5\,B & 7.0\,d & fine (16\,tokens) & global & \tyes \\
    \traceb (this work) & 515.8\,K & 9.7\,B & 130.9\,d & fine (16\,tokens) & global & \tyes \\
    \bottomrule
  \end{tabular}
  \begin{minipage}{\linewidth}
    \vspace{3pt}
    \footnotesize
    $^{\dagger}$Records are LLM interactions from 43 developers' individual sessions (1.4\,K rounds/day over eight months), not a continuous production service stream.

    $^{*}$The dataset comprises four independent traces, each spanning a two-hour window with its own localized clock.
  \end{minipage}
\end{table*}

To understand the unique character of LLM prefix cache, we begin by evaluating
how state-of-the-art caching algorithms perform on production traces.
We first establish two practical cache-size regimes for evaluation (\S\ref{sec:algorithms:regimes}),
and then directly compare fourteen online eviction algorithms against the optimal Belady's oracle
across two production serving traces (\S\ref{sec:algorithms:evaluation}).
Our results demonstrate that most complex algorithms yield little benefit over
a simple recency-based baseline (LRU), and in some cases, actively degrade
performance.

\begin{figure*}[t]
  \centering
  \includegraphics{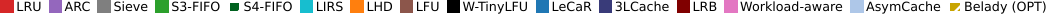}\\[1pt]
  \begin{subfigure}{0.49\linewidth}
    \centering
    \includegraphics[width=\linewidth]{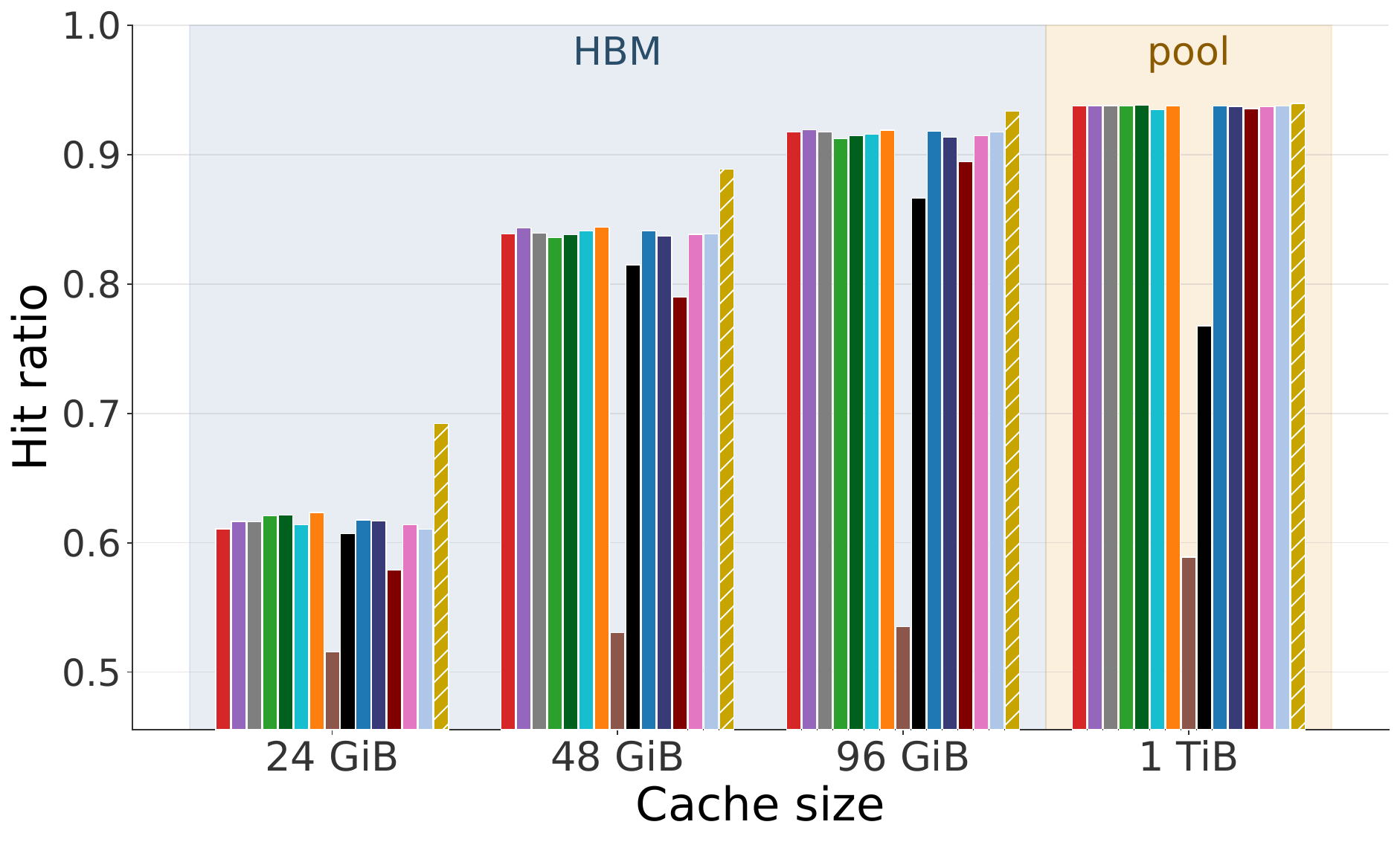}
    \vspace{-1.6em}
    \caption{\tracea.}
    \label{fig:bars-tracea}
  \end{subfigure}
  \hfill
  \begin{subfigure}{0.49\linewidth}
    \centering
    \includegraphics[width=\linewidth]{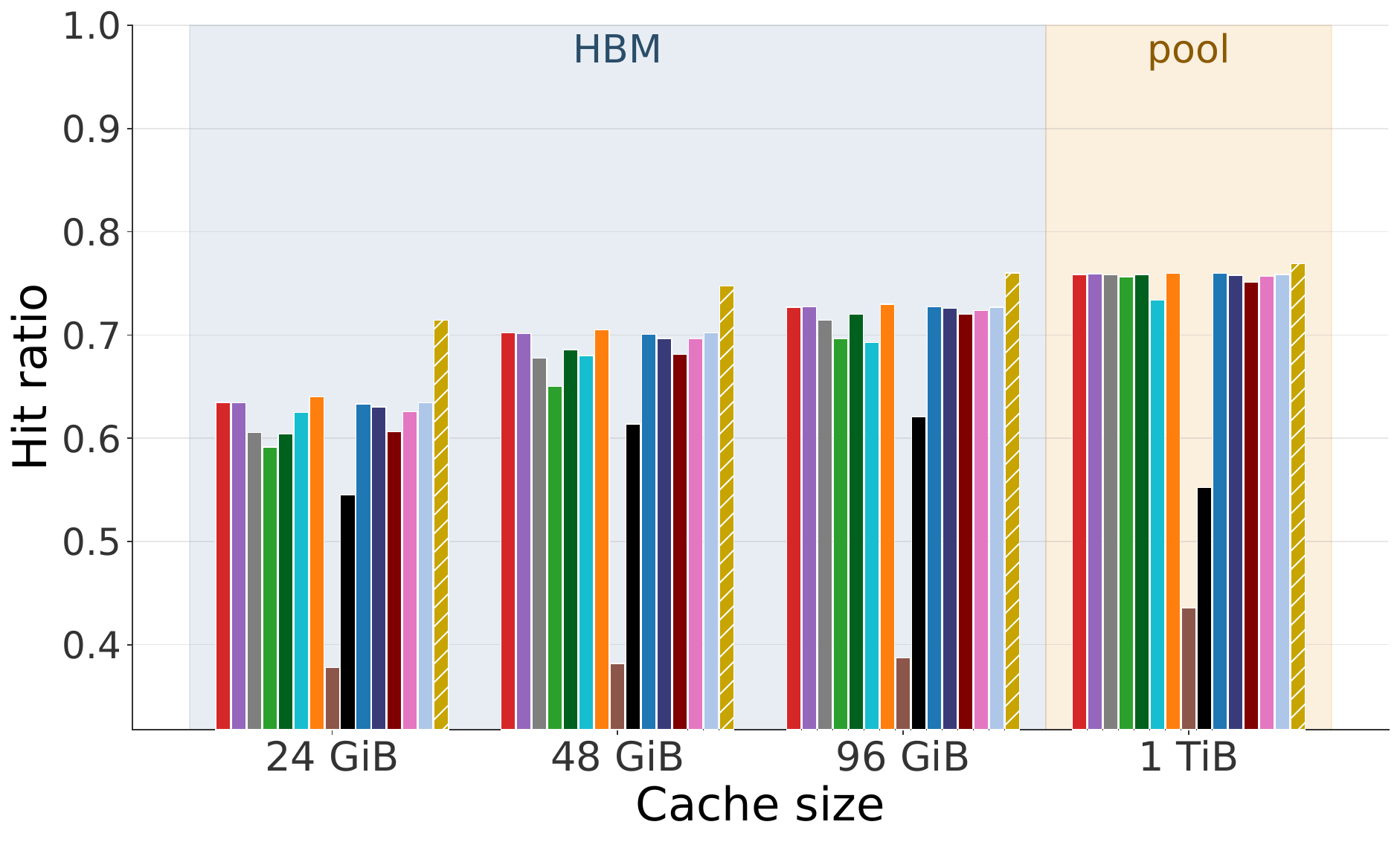}
    \vspace{-1.6em}
    \caption{\traceb.}
    \label{fig:bars-traceb}
  \end{subfigure}
  \vspace{-0.9em}
  \caption{Block hit ratio on the two production traces for fourteen eviction algorithms and offline optimum (Belady). 
  Across different memory capacities, none of the state-of-the-art algorithms improve over LRU.}
  \label{fig:bars-traces}
  \vspace{-0.5em}
\end{figure*}

\subsection{Traces and Simulator}
\label{sec:algorithms:regimes}

\noindent\textbf{Traces.}
We conduct our evaluation using two production traces from two different public LLM inference
services: \tracea~\cite{freeinference} and \traceb~\cite{chutes}.
\tracea is heavily dominated by agentic workloads with some non-agentic API usage,
while \traceb features a broader mix that includes multi-turn human conversations
alongside agentic sessions and non-agentic API usage.
To align with practical serving constraints, all evaluations use a 16-token block size and the \model~\cite{qwen3technicalreport} tokenizer, capping sequences at its 256k-token maximum context window (4.3\% skipped in \tracea and 0.03\% in \traceb).
Table~\ref{tab:traces} summarizes the key characteristics of each trace.
Table~\ref{tab:trace-compare} compares existing public datasets with the two traces introduced in this work. To accurately simulate prefix-cache eviction, a trace must provide fine-grained block identity scoped across sessions and contain multi-turn agentic traffic. Existing public traces lack at least one of these critical properties. The closest alternatives are the Qwen Bailian traces, which we also evaluate (\S\ref{sec:analysis:more}). While they offer 16-token block hashes, their session information reveals very little agentic workload and a much lower reuse ratio.

\begin{table*}[t]
  \caption{We compare three different caching workloads in this paper: LLM serving prefix cache, web cache (CDN), and block cache.}
  \vspace{-0.8em}
  \label{tab:workloads}
  \small
  \begin{tabular}{@{}ll rrr >{\raggedright\arraybackslash}p{6.75cm}@{}}
    \toprule
    Class & Trace & Requests & Distinct items & Span
      & Description \\
    \midrule
    Prefix cache & \tracea & 327.5\,K & 39.6\,M blocks & 7.0\,d
      & One week of a public LLM inference service. \\
    Web & Wikipedia 2019 text~\cite{cachemon} & 207.6\,M & 18.4\,M objects & 21.0\,d
      & Wikipedia's CDN request log for article text. \\
    Block & CloudPhysics w01~\cite{waldspurger2015shards,cachemon} & 215.8\,M
      & 122.8\,M objects & 6.3\,d
      & Disk I/O of one VM volume from a production fleet. \\
    \bottomrule
  \end{tabular}
  \vspace{-0.8em}
\end{table*}

\begin{table*}[t]
  \centering
  \caption{Summary of the characteristics that separate the prefix-cache
  workload from the web and block workloads. Each characteristic carries the
  subsection that measures it.}
  \vspace{-0.5em}
  \label{tab:workload-comparison}
  \small
  \begin{tabular}{@{}llll@{}}
    \toprule
    Characteristic & Prefix cache & Web & Block \\
    \midrule
    Lifetime (\S\ref{sec:analysis:lifetime}) & Minutes, bounded by the session
      & Days with a persistent hot core & Days without hot core \\
    Reuse interval (\S\ref{sec:analysis:reuse}) & Short, low variance & Long, high variance
      & Long, paced by periodic jobs \\
    Frequency (\S\ref{sec:analysis:frequency}) & Wide spread; tracks session progress
      & Zipfian; a few get most accesses & Few accesses; scans dominate \\
    Miss penalty (\S\ref{sec:analysis:miss-penalty}) & Grows with block depth & One uniform fetch
      & One uniform fetch \\
    Session footprint (\S\ref{sec:analysis:footprint}) & Heavy-tailed per session
      & No session structure & No session structure \\
    \bottomrule
  \end{tabular}
  \vspace{-0.8em}
\end{table*}

\noindent\textbf{Simulator.} 
To evaluate caching algorithms over these traces, standard block-level simulators like libCacheSim~\cite{libcachesim} are insufficient because they ignore the request-level residency restriction, which requires that all blocks for a request must reside in memory simultaneously during its prefill. To faithfully replicate real-world engine behavior, we developed \sys, a custom, high-performance C++ simulator based on libCacheSim. \sys proactively evicts blocks to ensure sufficient space is available to accommodate an incoming request's entire prefix at once, and its evaluated hit ratios are very close to those of the native vLLM~\cite{kwon2023efficient} engine. Compared to the vLLM's python implementation, this simulator yields up to 160$\times$ speedup in evaluation.

\noindent\textbf{Cache-size regimes.}
While memory per block varies by model, block-level eviction dynamics remain
consistent and the required capacity simply scales accordingly.
We evaluate eviction algorithms across two distinct capacity regimes,
summarized in Table~\ref{tab:cache-points}.
First, the \emph{per-replica HBM} regime models the constrained memory available directly on accelerators, ranging from 24\,GiB to 120\,GiB.
Scaling below 24\,GiB is impractical, as the residency constraint would severely restrict the maximum supported context length.
Second, the \emph{global DRAM/SSD memory pool} regime models massive disaggregated storage, spanning from 0.25\,TiB to 12\,TiB.

\subsection{Evaluation of Eviction algorithms}
\label{sec:algorithms:evaluation}

With our traces and simulator established, we sweep both cache-size regimes to evaluate how various caching mechanisms perform under a uniform cost model.
We report hit and compute-savings ratios as fractions, and differences between them in percentage points.

Figure~\ref{fig:bars-traces} shows the hit ratios of 14 online eviction algorithms across both production traces.
We group these algorithms by their primary design principle:
\emph{Recency}, acting as our baseline (LRU);
\emph{Quick demotion}, which aggressively evicts unproven blocks (ARC~\cite{megiddo2003arc}, Sieve~\cite{zhang2024sieve}, S3-FIFO~\cite{yang2023fifo}, S4-FIFO~\cite{xia2026s4fifo}, LIRS~\cite{jiang2002lirs});
\emph{Analytic modeling}, using fitted mathematical models (LHD~\cite{beckmann2018lhd});
\emph{Frequency} (LFU, W-TinyLFU~\cite{einziger2017tinylfu});
\emph{Learned} algorithms (LeCaR~\cite{vietri2018lecar}, LRB~\cite{song2020learning}, 3LCache~\cite{zhou20253lcache});
and recent \emph{domain-specific} prefix caching algorithms (Workload-aware~\cite{wang2025kvcache}, AsymCache~\cite{shi2026asymcache}).
Finally, we use Belady's optimum~\cite{belady1966study} to establish an \emph{offline ceiling} for comparison.

Our evaluation reveals that LRU remains remarkably robust
across prefix caching workloads, while more complex designs yield little benefit,
or even degrade performance.
Rather than outperforming this simple baseline, the advanced algorithms
fall into three behavioral groups.
First, a large majority of algorithms, including ARC, LHD, LeCaR, 3LCache, LRB,
Workload-aware and AsymCache, perform very similar, or slightly worse to LRU.
Despite their sophisticated designs, they merely reproduce
LRU's hit ratio without providing any meaningful advantage.
Second, quick demotion algorithms (S3-FIFO, S4-FIFO, Sieve, LIRS)
prove highly trace-dependent, offering marginal gains on some workloads
but regressing by several points on others.
Third, frequency-based algorithms, including LFU and W-TinyLFU,
completely collapse, trailing far behind.
Ultimately, there remains a significant headroom: Belady's offline oracle still sits
far above the best online algorithm. No current ``fancy''
mechanism fundamentally solves the eviction problem for prefix caching.

\section{What Makes Prefix-Cache Different}
\label{sec:analysis}

Traditional caching algorithms target web CDNs and block I/O, where objects are drawn from persistent global pools.
Prefix caching~\cite{kwon2023efficient,zheng2024sglang}, however, differs fundamentally.
Instead of drawing from static pools, its blocks accumulate dynamically through conversations and tool executions, creating distinct access behaviors.
To illustrate this, we compare our prefix-cache workload (\tracea from \S\ref{sec:algorithms}) with two canonical traces representing the primary workload classes in eviction-algorithm literature~\cite{cherkasova1998improving,beckmann2018lhd,yang2023fifo}: Wikipedia~\cite{cachemon} (web CDN) and CloudPhysics~\cite{waldspurger2015shards,cachemon} (production block I/O).
Table~\ref{tab:workloads} summarizes these three workloads.

We treat each KV block~\cite{vaswani2017attention} as a distinct cache object.
The following subsections detail the key characteristics distinguishing these workloads (summarized in Table~\ref{tab:workload-comparison}) and identify their root causes.

\begin{figure*}[t!]
    \centering
    \begin{subfigure}[t]{0.30\textwidth}
        \centering
        \includegraphics[width=\linewidth]{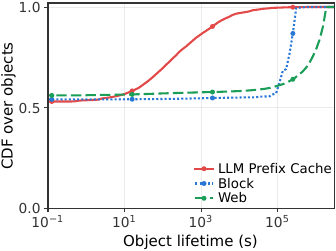}
        \caption{Object lifetime}
        \label{fig:lifetime-cdf}
    \end{subfigure}
    \hfill
    \begin{subfigure}[t]{0.30\textwidth}
        \centering
        \includegraphics[width=\linewidth]{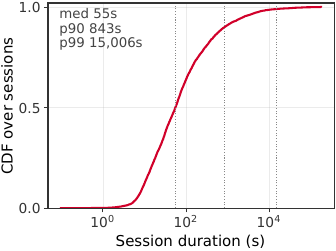}
        \caption{Session duration}
        \label{fig:session-duration}
    \end{subfigure}
    \hfill
    \begin{subfigure}[t]{0.30\textwidth}
        \centering
        \includegraphics[width=\linewidth]{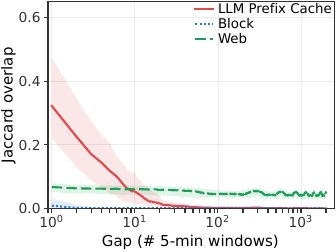}
        \caption{Similarity between consecutive working sets. }
        \label{fig:working-set-overlap}
    \end{subfigure}
    \vspace{-0.6em}
    \caption{(a) Prefix-cache blocks are short-lived, with lifetimes up to 100$\times$ shorter than in traditional caches. (b) Sessions are short: P90 is 14 minutes, while the median is under 1 minute. (c) Working set similarity is high at very short intervals \textit{due to short reuse} but rapidly falls to zero \textit{due to short lifetime}.\vspace{-0.8em}}
    \label{fig:prefix-cache-churn}
    \vspace{-0.5em}
\end{figure*}

\subsection{Session-Bounded Lifetimes}
\label{sec:analysis:lifetime}

Prefix-cache objects exhibit drastically shorter lifespans than those in traditional caching workloads.
Figure~\ref{fig:lifetime-cdf} illustrates this contrast: while traditional objects drawn from persistent pools remain active for days, \textit{most reused prefix-cache objects exhibit lifetimes under 30 minutes}.\footnote{While recent research has demonstrated long-running agents~\cite{li2026lhtbench,li2026agencybench,xu2026autolab}, most of the agents in production solve simpler tasks and run shorter.}
Because these blocks are generated by conversations, their utility is usually bounded by the originating session's lifespan.
Figure~\ref{fig:session-duration} further illustrates that
multi-turn sessions are predominantly short-lived
(median 55\,s, 99th percentile ${\sim}4$\,hours),
imposing a natural ceiling on the lifespans of most blocks.
Notably, the proportion of one-hit objects (lifetime of zero in Figure~\ref{fig:lifetime-cdf}) is similar across all three workloads (53.0\%--55.7\%).

Across the workload, prefix-cache blocks naturally divide into three categories with
distinct lifecycle dynamics:

\pgheading{System prompts} This category consists of system-role blocks appearing in
at least one multi-turn sessions. Despite comprising only 4.7\% of distinct blocks,
their cross-session reuse drives 18.4\% of all accesses.
They routinely outlive individual conversations,
persisting in the cache for days (median lifetime 291\,s, 99th percentile ${\sim}3.6$\,days).

\pgheading{Multi-turn session history} This category includes all non-system
blocks that appear in at least one multi-turn session.
Representing session-specific state like conversational turns and tool outputs,
these blocks constitute the primary driver of prefix reuse.
They make up 37.6\% of distinct blocks and generate an overwhelming
70.2\% of all cache accesses.
Because they are bound to specific conversations,
their lifetimes are strictly confined to their originating session
(median 219\,s).

\pgheading{Single-turn request prompts} Prompts from single-turn requests
form the largest fraction of unique blocks (57.8\%) but contribute
only 11.4\% of total accesses.
The vast majority exhibit virtually no reuse: 84.9\% of these blocks are
accessed exactly once.
The remaining 15.1\% of single-turn blocks correspond to shared prefixes
across different independent requests (such as common task templates).

This session-bounding also fundamentally changes working set churn.
Figure~\ref{fig:working-set-overlap} shows the Jaccard overlap between two 5-minute access windows over time. Web workloads maintain a persistent ${\sim}5\%$ globally popular core, while the block trace exhibits near-zero overlap.
The prefix-cache working set, however, completely lacks a persistent core: while adjacent windows share over 30\% of their objects, this overlap decays to near zero beyond a two-hour gap as sessions conclude.

This rapid churn and strict session-bounding have profound implications for cache sizing and eviction.
The working set size is straightforward to estimate---roughly the block arrival rate multiplied by the expected session duration.
As shown in Figure~\ref{fig:bars-traces},the 96\,GiB hit ratio is very high despite caching capacity under 0.2\% of unique blocks.
More critically, dedicating capacity to retain older blocks yields no benefit, since overlap vanishes after two hours.

\findingbox{Prefix-cache working sets churn rapidly: blocks have short lifetimes and lack the persistent hot set common in traditional cache workloads. As a result, most workloads do not need a large cache size (relative to unique data).}

\begin{figure}[t]
    \centering
    \begin{subfigure}[t]{0.48\linewidth}
        \centering
        \includegraphics[width=\linewidth]{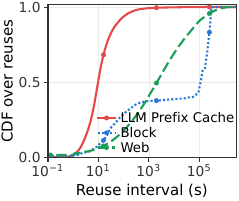}
        \vspace{-1.5em}
        \caption{Block reuse Interval}
        \label{fig:reuse-interval}
    \end{subfigure}
    \hfill
    \begin{subfigure}[t]{0.48\linewidth}
        \centering
        \includegraphics[width=\linewidth]{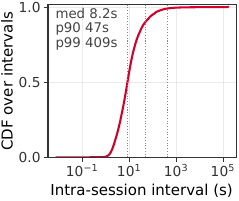}
        \vspace{-1.5em}
        \caption{Request inter-arrival}
        \label{fig:session-interval-dist}
    \end{subfigure}
    \vspace{-0.8em}
    \caption{(a) Prefix-cache reuse intervals are short with low variance. (b) Request inter-interval time within a session is short.}
    \label{fig:reuse-session-interval}
\end{figure}

\begin{figure*}[t!]
    \centering
    \begin{subfigure}[t]{0.32\textwidth}
        \centering
        \includegraphics[width=\linewidth]{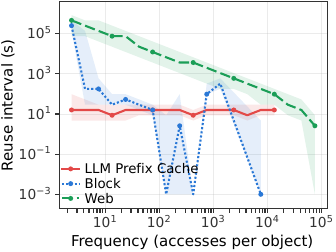}
        \caption{Reuse interval vs.\ frequency. Lines show the median and bands the P25--P75 range.}
        \label{fig:next-access-freq}
    \end{subfigure}
    \hfill
    \begin{subfigure}[t]{0.32\textwidth}
        \centering
        \includegraphics[width=\linewidth]{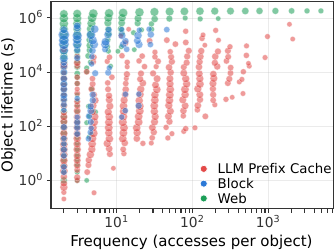}
        \caption{Object lifetime vs.\ frequency. Bubble area indicates the number of objects in each bin.}
        \label{fig:popular-shortlife}
    \end{subfigure}
    \hfill
    \begin{subfigure}[t]{0.32\textwidth}
        \centering
        \includegraphics[width=\linewidth]{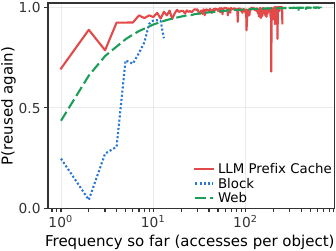}
        \caption{Probability of further reuse vs.\ current frequency.}
        \label{fig:nextfreq-cdf}
    \end{subfigure}
    \vspace{-0.8em}
    \caption{Frequency is not a good indicator for LLM prefix cache. (a) Reuse intervals are largely independent of block popularity in the LLM prefix cache. (b) Block lifetime increases with access frequency and shows a wide spectrum. (c) Frequency is a poor predictor of future reuse. }
    \label{fig:frequency-analysis}
\end{figure*}

\begin{figure}[t]
\begin{subfigure}[t]{0.48\linewidth}
  \centering
  \includegraphics[width=\linewidth]{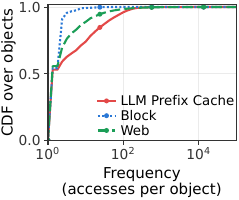}
  \caption{Frequency.}
  \label{fig:reuse-count-cdf}
\end{subfigure}
\hfill
\begin{subfigure}[t]{0.48\linewidth}
  \centering
  \includegraphics[width=\linewidth]{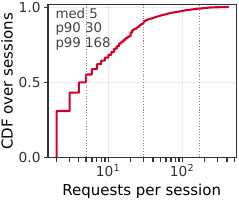}
  \caption{Requests per multi-turn session.}
  \label{fig:session-requests-dist}
\end{subfigure}
\vspace{-0.8em}
\caption{(a) Prefix-cache blocks are reused more frequently than in web and storage block caches. (b) Requests per session show a wide spectrum.}
\end{figure}

\subsection{Recency Provides a Key Signal}
\label{sec:analysis:reuse}

Prefix-cache reuse intervals are remarkably short and concentrated, contrasting sharply with the long, high-variance intervals of traditional workloads,  as shown in Figure~\ref{fig:reuse-interval}.
Web interarrival times reflect independent users, producing smooth curves with a wide interquartile range (P25--P75).
Block traces mix diverse behaviors like periodic scans, causing sharp modal spikes.
Prefix caching, however, is dominated by rapid intra-session interactions.
Figure~\ref{fig:session-interval-dist} highlights this extreme recency.
Driven by quick tool executions and conversational turns, the median intra-session interval is a mere 8.2\,s, and 99.7\% of all gaps fall under a 22-minute reuse window.

While the median reuse interval is short across all block categories (10.0\,s
for system prompts, 11.2\,s for multi-turn history), they do exhibit a heavy
tail, with the 99th percentile reaching 1{,}000\,s and 562\,s respectively. This
tail is caused by extended human interaction and prolonged tool executions,
which can delay the next request up to several days.

Despite this heavy tail, because the overwhelming majority of reuses are concentrated in extremely short intervals, recency remains the dominant signal for whether a block should be retained.
When a block goes idle for a long period, it should be aggressively evicted regardless of its past popularity.
This is because a long-idle block likely belongs to a working set that has already churned away, meaning it will probably not be reused.
Furthermore, even if the block does fall into the heavy tail and is eventually reused, the capacity cost of pinning it in memory throughout a prolonged idle duration far outweighs the benefit of a single cache hit.
This explains why simple recency-based eviction algorithms perform so well in our initial study of existing algorithms (\S\ref{sec:algorithms:evaluation}).

\findingbox{Eviction should prioritize recency: reuse is concentrated within short intervals, long-idle blocks should be reclaimed aggressively, as they are unlikely to be reused and increasingly costly to retain.}

\subsection{Frequency Tracks Session Progress}
\label{sec:analysis:frequency}

Prefix-cache objects exhibit a vastly wider spread of frequency than traditional workloads.
As demonstrated in Figure~\ref{fig:reuse-count-cdf}, while web frequencies plateau early due to Zipfian popularity and block objects see minimal reuse, prefix-cache frequency fundamentally tracks session progress.
As sessions advance, their history blocks are repeatedly re-read.
Because a block's reuse count depends on both the overall session length and its insertion turn, reuses span from a handful to hundreds of times.
Figure~\ref{fig:session-requests-dist} shows that the number of turns per session is exceptionally large and heavy-tailed, reaching up to 412 requests.
This massive variation in session lengths drives extreme frequency variance across the cache, with system prompts and multi-turn history blocks reaching median accesses of 15 and 12 respectively.
Such prolonged interactions also cause immense context growth, making partial eviction highly valuable for preserving the active context.

Unlike traditional workloads, accumulated frequency in prefix caching provides almost no signal about the timing of the next request.
As shown in Figure~\ref{fig:next-access-freq}, web and block traces exhibit a downward trend where higher popularity translates to shorter reuse intervals due to concurrent user requests.
In contrast, the prefix-cache reuse interval remains completely flat across all frequency bins, demonstrating that accesses arrive at the stable, sequential pace of individual sessions.
Consequently, prefix-cache object lifetime increases as frequency grows, as demonstrated in Figure~\ref{fig:popular-shortlife}.
Web and block lifetimes, however, remain decoupled from frequency, since objects drawn from a global pool stay active indefinitely with little reuse.
Because frequency fails to predict shorter reuse intervals, it proves highly ineffective for ranking eviction candidates.
This fundamental disconnect explains why frequency-heavy algorithms suffer such poor performance in our initial algorithm comparison (\S\ref{sec:algorithms:evaluation}).

However, early frequency is an excellent predictor for \emph{whether} a block will be reused at all.
As shown in Figure~\ref{fig:nextfreq-cdf}, prefix caching exhibits a uniquely sharp increase in future reuse probability during the first few hits.
Securing just a second hit jumps this probability from 69.4\% to 88.7\%, cleanly separating single-turn requests from established multi-turn sessions.
For workloads flooded with one-hit wonders, this implicit classification makes early frequency a helpful admission filter to discard transient prompts.
We explore this conditional benefit further in \S\ref{sec:designs:qd}, demonstrating how quick demotion specifically aids the Qwen To-B trace~\cite{wang2025kvcache}.

\findingbox{Because session accesses arrive at a steady pace, frequency fails to predict reuse intervals and should not be used as an eviction ranking metric. However, for workloads dominated by one-hit wonders, early frequency can serve as a helpful admission filter.}

\subsection{Position-Dependent Miss Costs}
\label{sec:analysis:miss-penalty}

\begin{figure}[t]
  \centering
  \includegraphics[width=0.96\linewidth]{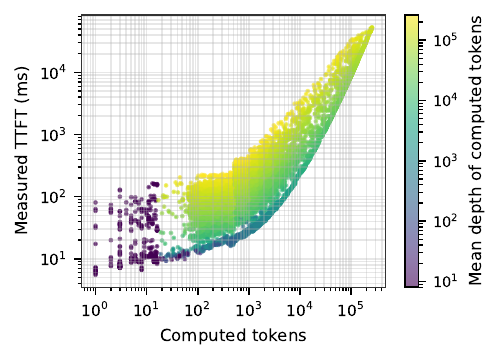}
  \vspace{-0.8em}
  \caption{Measured TTFT against the number of tokens computed per request,
  over 10{,}000 \tracea requests replayed on
  vLLM~\cite{kwon2023efficient} with
  \model~\cite{qwen3technicalreport} under LRU at a 48\,GiB cache on a single
  NVIDIA H200.
  Color indicates the mean depth (position in the prompt) of the computed
  tokens. Similar token counts can incur substantially different TTFT.}
  \label{fig:ttft-vs-computed}
\end{figure}

\begin{figure}[t]
  \centering
  \includegraphics[width=0.95\linewidth]{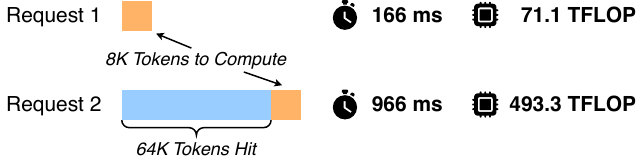}
  \vspace{-0.8em}
  \caption{Depth-dependent recomputation cost on
  \model~\cite{qwen3technicalreport}, measured on NVIDIA H200. Both
  requests compute 8K
  tokens, but the second must attend to an additional 64K cached tokens,
  increasing computation and TTFT.}
  \label{fig:computation-aware}
\end{figure}

\begin{figure}[t]
  \centering
  \includegraphics[width=0.8\linewidth]{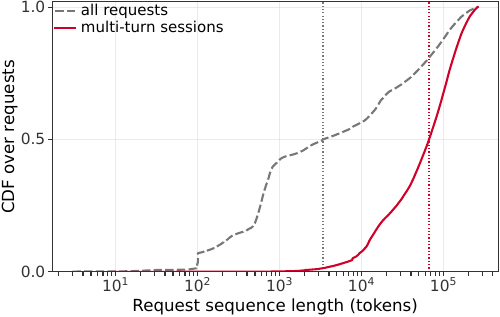}
  \vspace{-0.5em}
  \caption{Distribution of request sequence lengths in tokens, over all
  requests and over the requests that belong to multi-turn sessions. Long
  requests both occupy substantial HBM and create deep blocks with high
  recomputation costs. Dotted lines mark the two medians, 3{,}386 and
  66{,}605 tokens.}
  \label{fig:seqlen-dist}
\end{figure}

Hit ratio alone does not determine the benefit of prefix caching.
Figure~\ref{fig:ttft-vs-computed} plots measured TTFT against the number of
tokens computed per request.
Even among requests computing a comparable number of tokens, TTFT varies
substantially. The color of each point shows that requests computing tokens
deeper in the prompt generally incur higher TTFT at a similar token count.
Thus, the absolute number of tokens left to compute is insufficient
to predict TTFT.

Figure~\ref{fig:computation-aware} illustrates why depth matters.
Both requests compute 8K tokens, but one starts from scratch
while the other follows a cached 64K-token prefix.
Each uncached token must attend to all preceding tokens in full-attention layers,
including those cached tokens.
Hence, the second request performs more attention work despite computing
the same number of tokens. For a fixed-size block, this attention work
grows with its depth, making misses on deeper blocks more expensive.

These cost differences matter because request lengths span a wide range
(Figure~\ref{fig:seqlen-dist}), reaching 111K tokens at the 90th percentile.
Such long contexts contain blocks with vastly different recomputation costs,
despite occupying identical cache space and contributing equally to hit ratio.

Eviction algorithms should therefore account for the computation saved by
each hit. When blocks have comparable expected reuse, retaining deeper,
more expensive blocks and evicting cheaper, shallow blocks can reduce
recomputation cost for the same number of hits. The separation of cache
residency from prefix matching (\S\ref{sec:background:prefix-caching}) makes
this possible by allowing shallow blocks to be recomputed while deeper
cached blocks are reused. We will evaluate compute-aware eviction and its
implementation tradeoffs in \S\ref{sec:designs:compute}.

\findingbox{Hit ratio alone does not determine TTFT because deeper blocks save more computation. Eviction should thus account for this depth-dependent cost.}

\subsection{Heavy-Tailed Session Footprints}
\label{sec:analysis:footprint}

\begin{figure}[t]
  \centering
  \begin{subfigure}[t]{0.48\linewidth}
    \centering
    \includegraphics[width=\linewidth]{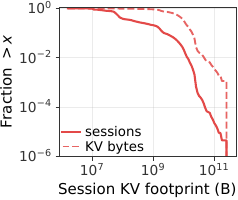}
    \caption{Footprint tail (CCDF), by sessions and KV bytes.}
    \label{fig:session-footprint-ccdf}
  \end{subfigure}
  \hfill
  \begin{subfigure}[t]{0.48\linewidth}
    \centering
    \includegraphics[width=\linewidth]{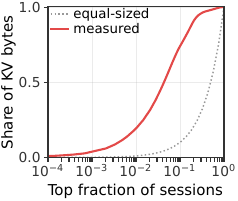}
    \caption{Concentration of KV bytes across sessions.}
    \label{fig:session-footprint-lorenz}
  \end{subfigure}
  \vspace{-0.5em}
  \caption{Per-session KV footprint of the prefix-cache trace.}
  \label{fig:session-footprint}
\end{figure}

\begin{figure*}[t]
  \centering
  \includegraphics[width=\linewidth]{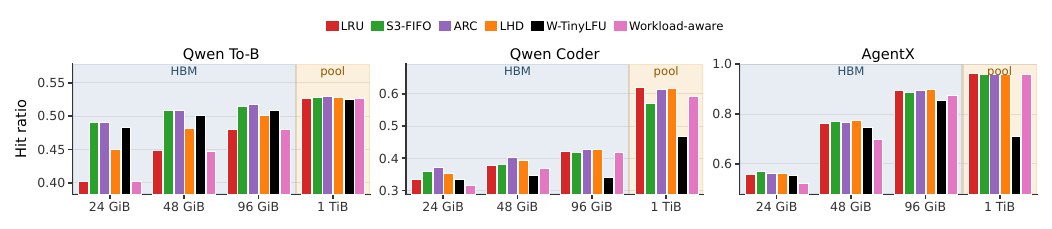}
  \vspace{-2em}
  \caption{Block hit ratio of six eviction algorithms at the four capacities of
  Table~\ref{tab:cache-points}, on two of the four Qwen Bailian
  traces~\cite{wang2025kvcache} and on AgentX, a replay of 393 Claude Code
  sessions~\cite{semianalysis2026agentx}.}
  \label{fig:bars8-qwen}
\end{figure*}

Session KV footprints span several orders of magnitude.
Figure~\ref{fig:session-footprint-ccdf} shows the median session accumulates just 45 blocks (0.07\,GiB),
whereas the 99th percentile reaches 8.3K blocks (12.2\,GiB) and the maximum hits 234\,GiB.
These massive footprints emerge as context accumulates across turns (\S\ref{sec:analysis:frequency}) and scales with the long requests shown in Figure~\ref{fig:seqlen-dist}, allowing a single session to easily overwhelm the HBM cache.

Consequently, KV memory concentrates intensely in the largest sessions.
The Lorenz curve in Figure~\ref{fig:session-footprint-lorenz} reveals the top 10\% of sessions hoard 76.2\% of all KV bytes, with the top 1\% alone consuming 20.3\%.
By partially trimming these heavy sessions, cache managers can reclaim substantial capacity without destroying the entire reusable context.

This extreme skew forces eviction granularity to adapt to capacity constraints.
When cache space is tight, treating a massive session as an indivisible unit forces a destructive all-or-nothing choice,
while fine-grained eviction gracefully avoids this by reclaiming selected blocks.
Conversely, in an expansive secondary pool,
session-level management becomes highly attractive for drastically reducing metadata overhead.
We quantify this tradeoff in \S\ref{sec:designs:granularity}.

\findingbox{Session KV footprints are heavy-tailed. Constrained caches require partial-session eviction, while larger pools may instead trade fine-grained eviction
for lower metadata overhead.}

\subsection{Findings Generalize Across Workloads}
\label{sec:analysis:more}

To test whether our findings generalize beyond \tracea, we extend the algorithm comparison from \S\ref{sec:algorithms:evaluation} to two production Qwen Bailian traces~\cite{wang2025kvcache} and AgentX~\cite{semianalysis2026agentx}. We evaluate a six-algorithm subset at four cache capacities in Figure~\ref{fig:bars8-qwen}, selecting canonical algorithms for each category.

AgentX replays 393 Claude Code sessions from the SemiAnalysis corpus~\cite{semianalysis2026agentx}. Because the corpus lacks absolute timestamps, we artificially schedule sessions to maintain a steady concurrency of 16, while preserving the exact timing between requests within each session. Because its hashes are session-scoped, cross-session sharing is modeled as zero, making the AgentX hit ratios a conservative lower bound.

The results confirm the core role of recency (\S\ref{sec:analysis:reuse}).
At each capacity, the best online algorithm is usually recency-based, and
by 1\,TiB, LRU is within 0.4 points of the best online algorithm on every trace.
Conversely, frequency-based W-TinyLFU consistently underperforms (\S\ref{sec:analysis:frequency}).

These three traces represent distinct serving workloads. Across them, the recency
conclusion is stable, but the capacity at which LRU catches up is
workload-dependent. The Qwen traces contain less reusable content than \tracea:
compulsory misses account for 34--54\% of accesses, compared with 6.0\% on \tracea.
Conversely, AgentX contains even more reusable content, with only 3.75\% compulsory misses.
Accordingly, the offline optimum at 1\,TiB is 0.534 (To-B), 0.664 (Coder), and 0.962 (AgentX).
At 24\,GiB, LRU does not lead on any trace; on Qwen To-B, for
example, S3-FIFO and ARC reach 0.491 versus LRU's 0.403. LRU catches up only at
60--462\,GiB, compared with 37\,GiB on \tracea. Thus, some workloads require
substantially more cache capacity before recency alone is sufficient. This
motivates the conditional use of quick demotion in
\S\ref{sec:designs:qd}. Appendix~\ref{sec:appendix:more-traces} presents line
plots across the full capacity range for the four Qwen traces, AgentX, and the two
traces evaluated in \S\ref{sec:algorithms}, each with the full algorithm set;
Appendix~\ref{sec:appendix:concsens} analyzes sensitivity to session
concurrency at fixed capacity.

\section{What Designs Can Be Helpful?}
\label{sec:designs}

The empirical policy behaviors observed in \S\ref{sec:algorithms} and the
design implications uncovered in \S\ref{sec:analysis} demonstrate that
while a standard recency-based policy like LRU provides a remarkably robust
baseline, no single caching algorithm universally wins across all deployment scenarios.
Because optimal performance is intimately tied to the interaction between 
workload characteristics and cache capacity, we do not propose a monolithic ``best'' policy. 
Instead, we explore a set of modular algorithmic techniques that can be selectively deployed to address 
the specific bottlenecks of prefix caching.

\textbf{LRU as a bedrock.}
Because recency is the primary predictor of future reuse (\S\ref{sec:analysis:reuse}), 
Least Recently Used (LRU) serves as the foundational baseline for our proposed designs. 
As shown in \S\ref{sec:algorithms}, plain LRU is remarkably robust, reliably converging 
to the optimal offline ceiling on every trace we measured while requiring the absolute 
minimum metadata overhead. Consequently, rather than reinventing the core eviction logic, 
we use LRU as the bedrock and introduce several targeted add-ons designed to handle 
specific edge cases and constrained environments where pure recency falls short.

\subsection{Quick Demotion}
\label{sec:designs:qd}

When capacity is severely constrained and the workload is flooded with single-turn
requests, recency alone can struggle to protect the active working set. In environments dominated by
one-hit wonders, \emph{quick demotion} strategies like S3-FIFO~\cite{yang2023fifo}
can be highly effective by rapidly filtering out blocks that will never be reused (\S\ref{sec:analysis:frequency}).
Figure~\ref{fig:bars8-qwen} (Qwen To-B panel) shows the value of this filtering:
at 24\,GiB, S3-FIFO achieves a 0.491 hit ratio compared to LRU's 0.403.

However, one-hit wonders typically belong to short, single-turn prompts,
meaning they occupy a small cache footprint compared to multi-turn sessions.
Therefore, quick demotion provides substantial benefits only under specific
extremes: when one-hit prompts account for such an overwhelming volume of
traffic that their cumulative footprint genuinely threatens to flood the cache.

\textbf{Takeaway.} Quick demotion strategies like S3-FIFO are effective, but should only be enabled when one-hit prompts account for an extremely high volume of traffic.

\subsection{Compute-awareness}
\label{sec:designs:compute}

\begin{figure}[t]
  \centering
  \includegraphics[width=\linewidth]{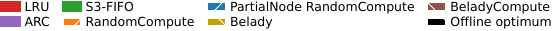}\\[2pt]
  \includegraphics[width=\linewidth]{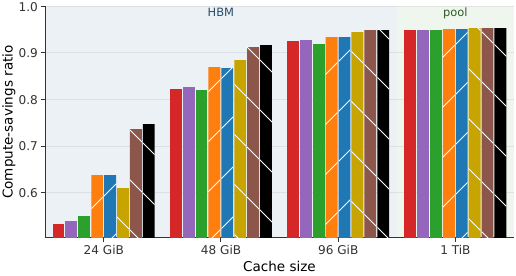}
  \caption{Compute-savings ratio on \tracea under the \model~\cite{qwen3technicalreport} cost model. BeladyCompute tightly approximates the ILP optimum, whereas standard Belady falls short. Compute-aware eviction substantially improves compute savings, and partial-node eviction preserves these gains without degradation.}
  \label{fig:compute-aware}
\end{figure}

Because a cache miss in LLM serving triggers recomputation, 
eviction decisions should ideally optimize for compute savings rather than pure hit counts, 
especially when the cost of recomputation varies widely across blocks. To evaluate this, we 
introduce the \emph{compute-savings ratio}, a metric that weights each cache hit by the 
computational cost avoided. Under this framework, the traditional hit ratio simply represents 
a constant cost model that regards all blocks as having the exact same cost. To capture real-world 
variance, we model compute intensity based on the FLOPs required for each block on 
Qwen3 Coder 30B. Figure~\ref{fig:compute-aware} illustrates the compute-savings 
ratio of various policies under this cost model. We additionally evaluate an extreme linear 
cost model where full attention strictly dominates, finding that all conclusions 
below hold with slightly wider margins (Appendix~\ref{sec:appendix:linear}).

\begin{figure}[t]
  \centering
  \includegraphics{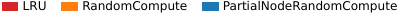}
  \begin{subfigure}[t]{0.48\linewidth}
    \centering
    \includegraphics[width=\linewidth]{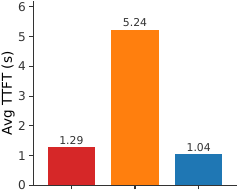}
    \vspace{-0.8em}
    \caption{Average TTFT.}
    \label{fig:e2e-ttft}
  \end{subfigure}
  \hfill
  \begin{subfigure}[t]{0.48\linewidth}
    \centering
    \includegraphics[width=\linewidth]{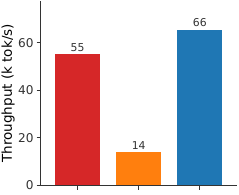}
    \vspace{-0.8em}
    \caption{Prefill throughput.}
    \label{fig:e2e-throughput}
  \end{subfigure}
  \caption{End-to-end performance on H200 with 48\,GiB cache replaying 10{,}000 requests from \tracea. TTFT~(a) and throughput~(b) were measured in separate runs. Partial-node compute-aware eviction yields substantial TTFT and throughput gains over LRU, whereas block-based eviction incurs severe degradation due to fragmentation.}
  \label{fig:e2e}
\end{figure}

\begin{figure*}[t]
  \centering
  \begin{subfigure}{0.48\linewidth}
    \centering
    \includegraphics[width=\linewidth]{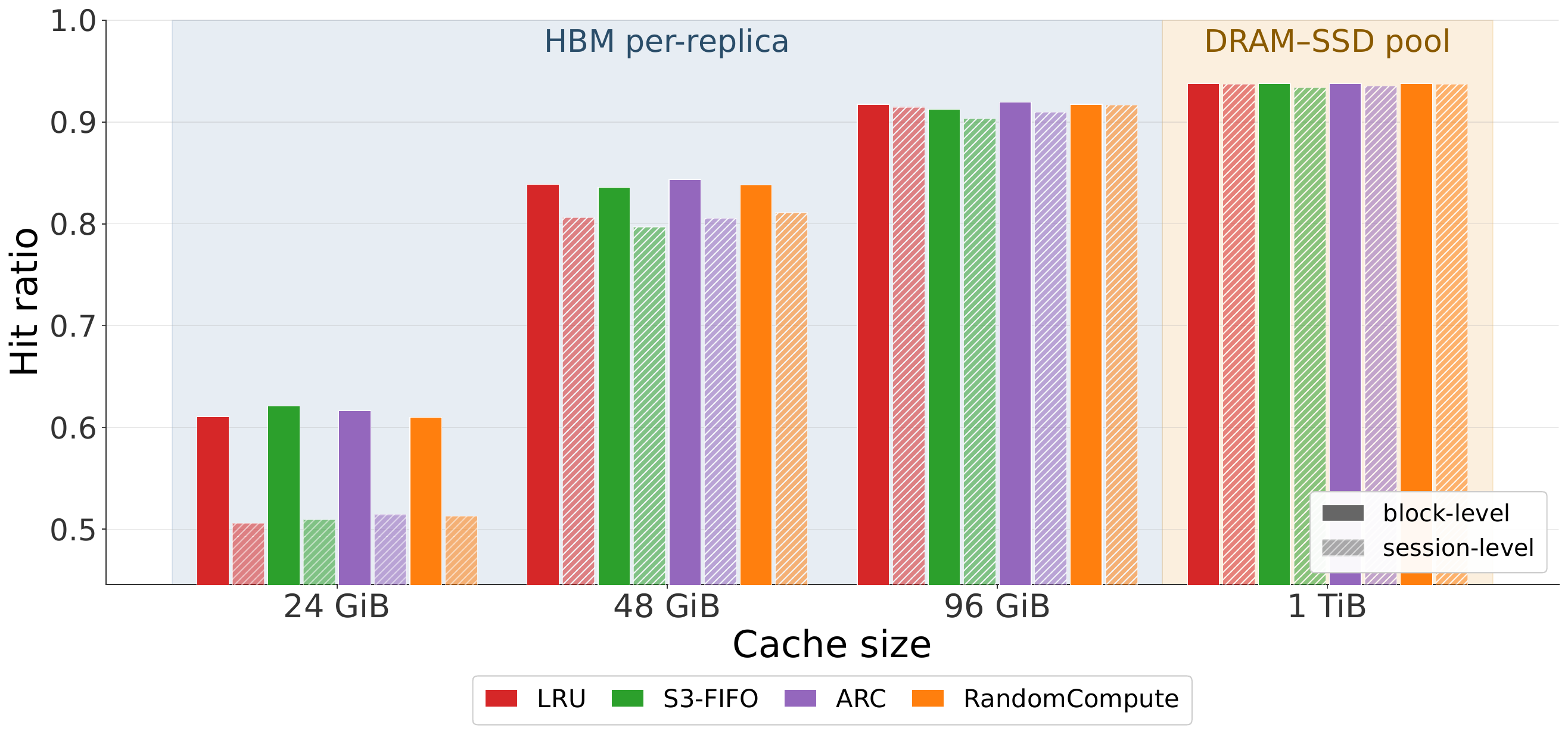}
    \caption{Hit ratio at four capacities.}
    \label{fig:svb-bars-constant}
  \end{subfigure}
  \hfill
  \begin{subfigure}{0.48\linewidth}
    \centering
    \includegraphics[width=\linewidth]{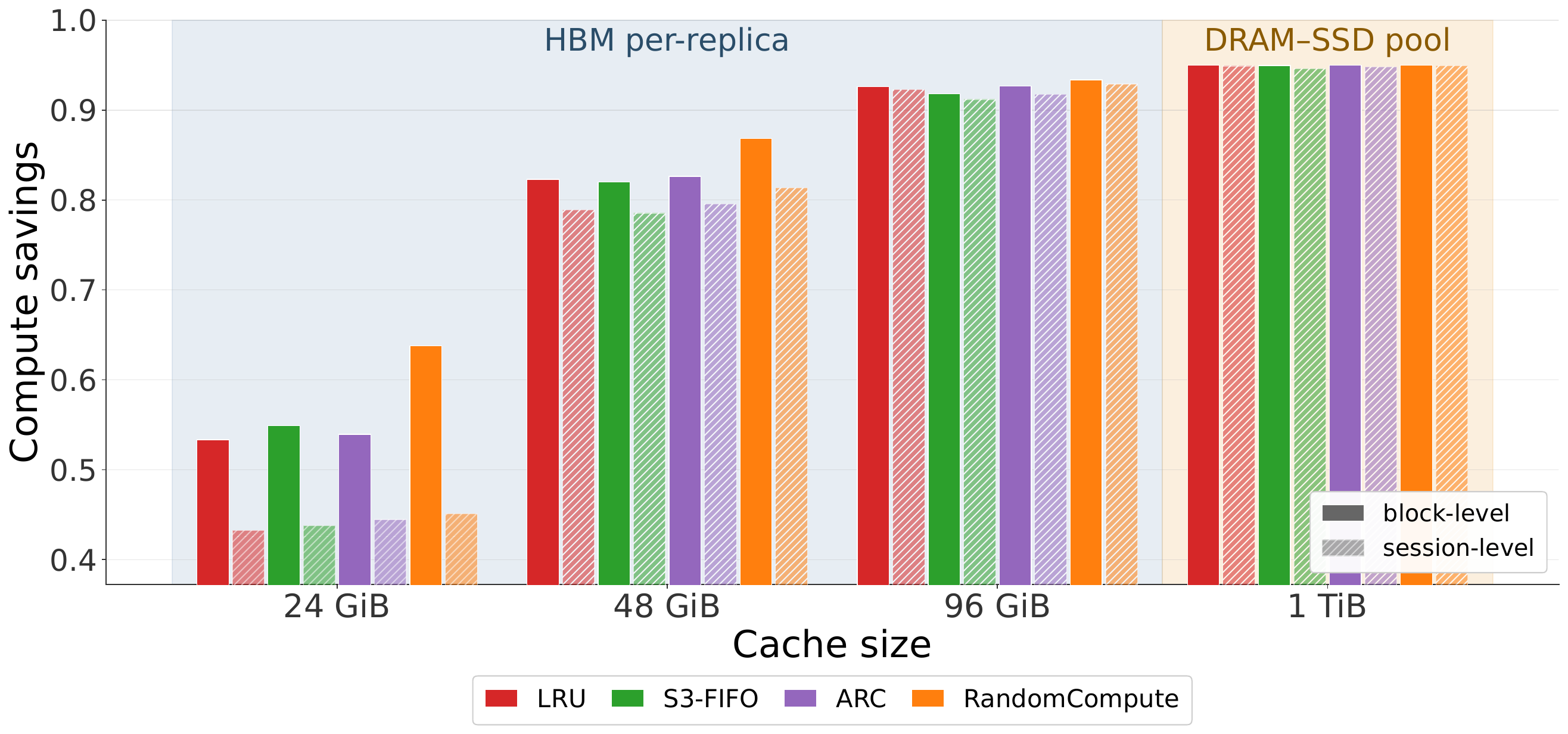}
    \caption{Compute savings at four capacities.}
    \label{fig:svb-bars-qwen}
  \end{subfigure}
  \vspace{-0.5em}
  \caption{Block-level (solid) vs. session-level (hatched) eviction on the prefix-cache trace. While session-level eviction matches block-level performance at large capacities, it degrades under tight constraints.}
  \label{fig:session-vs-block}
\end{figure*}

To establish a theoretical optimum for this metric, we frame the caching process as an 
Integer Linear Programming (ILP) problem, using reuse intervals as variables and capacity 
as a constraint. Because solving the ILP is computationally expensive, we also introduce 
\emph{BeladyCompute}: an efficient offline algorithm inspired by
BeladySize~\cite{libcachesim}. For a given block $i$, 
BeladyCompute evaluates an eviction score:
\vspace{-0.3em}
\[
    \text{Score}(i) = \text{ComputeIntensity}(i) \times \text{TimeUntilNextAccess}(i) \\\vspace{-0.3em}
\]
and evicts the block with the highest score. As shown in Figure~\ref{fig:compute-aware}, BeladyCompute 
tracks the offline ILP optimum exceptionally well across all constrained cache sizes---falling 
just 1.03 points below the optimum at 24\,GiB, 0.41 points at 48\,GiB, and a 0.08 points at 96\,GiB. This proves it is a robust and efficient approximation of the absolute theoretical limit.

These offline oracles offer an easy diagnostic for whether a workload deserves 
compute-awareness. The gap in compute-savings ratio between the hit-optimal Belady and 
cost-aware BeladyCompute bounds the potential gain. On \tracea, this gap 
is massive (+12.6 points at 24\,GiB), justifying the design effort. Conversely, on the 
other five traces, the gap is negligible ($\le$ 1.1 points), meaning standard hit-optimal 
evictions is good enough. Thus, running both oracles on a trace sample can decisively evaluate the necessity of compute-awareness before designing any online policy. Because this cost profile depends primarily on the model and the proportion of long sequences, the diagnostic only needs to be repeated when these factors change.

For practical online environments, we design \emph{RandomCompute}, a compute-aware algorithm 
that targets cheap, shallow, and long-idle blocks for eviction. It achieves this by randomly 
sampling a subset of blocks in the cache and evicting the one with the lowest score, calculated 
for each sampled block $i$ as:
\vspace{-0.3em}
\[
    \text{Score}(i) = \frac{\text{ComputeIntensity}(i)}{\text{TimeSinceLastAccess}(i)}\\\vspace{-0.3em}
\]
In long-context scenarios with high variance 
in block computation costs, this strategy yields massive efficiency gains. At a constrained 24\,GiB 
cache under this cost model, RandomCompute achieves a 0.638 compute-savings ratio, 
outperforming LRU by 10.4 points and even surpassing the standard hit-optimal Belady
oracle~\cite{belady1966study}.

However, as a sample-based algorithm, RandomCompute introduces a severe problem: it 
heavily fragments the prefix, leaving behind numerous ``holes''\footnote{A hole is a contiguous
run of blocks a request has to recompute.} that creates massive overhead
of the attention matrix computation~\cite{vaswani2017attention}. RandomCompute leaves
an average of 9.21 holes per request, with a long tail reaching hundreds of holes.
Appendix~\ref{sec:appendix:holes} reports the full distributions. To mitigate 
this fragmentation, we introduce \emph{partial-node} eviction, which drops contiguous chunks 
of blocks from the beginning of a node rather than scattered individual blocks. Preferably, this chunk 
size should align with the system's contiguous batching token budget (if applicable).
This alignment ensures that hole-filling perfectly utilizes the budget while minimizing any reduction in compute savings. 
Furthermore, because there are no actual computational dependencies between missing segments, 
the prefill for different holes can be batched and executed entirely in parallel. By packing the same 
recomputation volume into fewer and longer holes, PartialNode RandomCompute slashes the 
average hole count by 8.7$\times$. This optimization effectively eliminates the parallelization 
bottleneck while sacrificing negligible compute savings at 24\,GiB, making it a highly practical design.

We validate these gains on a real vLLM engine serving \model~\cite{qwen3technicalreport} with a 48\,GiB cache (Figure~\ref{fig:e2e}). Against LRU, our partial-node compute-aware design cuts average Time-To-First-Token (TTFT) by 19.9\% (1.04\,s vs.\ 1.29\,s) and raises prefill throughput by 18.8\% (65.6K vs.\ 55.2K tokens/s). Fragmentation penalties render the original block-granular RandomCompute unusable: it incurs 4.0$\times$ higher TTFT and 3.9$\times$ lower throughput than LRU. Mechanistically, our policy trades cheap shallow hits to protect expensive deep ones: while LRU retains a slightly better median TTFT (112\,ms vs.\ 171\,ms), our design decisively slashes the 99th-percentile tail from 26.0\,s down to 16.8\,s.

\textbf{Takeaway.} When sequence lengths are long and recomputation costs vary significantly,
compute-aware eviction delivers massive efficiency gains.
However, practical online implementations must evict blocks in contiguous chunks
(e.g., partial-node eviction) to avoid severe memory fragmentation that would otherwise
cripple parallel attention computation.

\subsection{Block-Level vs. Session-Level Eviction}
\label{sec:designs:granularity}

When managing secondary KV cache pools~\cite{qin2025mooncake,xie2025strata},
maintaining metadata for millions of 
individual blocks incurs a significant overhead.
For example, tracking a 12\,TiB pool requires indexing 8.4~million 16-token blocks,
so systems like Mooncake~\cite{qin2025mooncake} and LMCache~\cite{liu2025lmcache} use
16--32$\times$ coarser chunks (256--512 tokens).
\emph{Session-level eviction} drastically reduces this overhead by evicting
entire sessions at once, but sacrifices the granular efficiency of partial eviction.

Figure~\ref{fig:session-vs-block} illustrates this fundamental tradeoff.
At highly constrained cache sizes (e.g., 24\,GiB), 
session-level eviction is disastrous: LRU loses 10.5 points of hit ratio, and the 
compute-aware RandomCompute policy loses a staggering 18.7 points of compute savings.
When a giant session monopolizes the cache, failing to partially trim it forces many valuable sessions out.
The penalty reduces but stays material at the 48\,GiB, where LRU still 
gives up 3.3 points of hit ratio and RandomCompute 5.5 points of compute savings. Since blocks are 
already the native management granularity in HBM, accepting this penalty offers no system benefit.
This severe penalty falls off quickly as capacity grows. In the large pool 
regime, the performance gap between session-level and block-level eviction 
becomes virtually unmeasurable. The only permanent drawback to session granularity is 
cross-session sharing: evicting an entire session inadvertently purges shared system prompts 
that other active sessions still need, though the vast capacity of the pool largely masks this effect.

\textbf{Takeaway.} Eviction granularity needs to match the memory tier. In the HBM regime, 
systems should strictly employ block-level eviction to maximize performance, which naturally align 
with native block management as well. Conversely, given the performance penalty vanishes in 
the large pool regime, designers can freely choose between block-level and 
session-level eviction based on metadata organization.

\subsection{Summary}

In summary, designing a cache for prefix-caching workloads requires an approach tailored to specific capacity constraints and workload characteristics. If the workload is flooded with an extremely high volume of short, single-turn prompts that threatens to overwhelm capacity, systems should enable quick demotion (e.g., S3-FIFO) to filter them out early. If the system serves models with heavy attention complexity and long sequences, designers should switch to a compute-aware eviction algorithm (paired with partial-node granularity) to prioritize retaining expensive blocks without causing fragmentation. When combined with quick demotion, this compute-awareness should only be applied to the main cache to prevent the admission queue from prematurely evicting shallow blocks. Finally, if managing a constrained HBM cache, strictly employ block-level eviction to maximize performance; conversely, if managing a massive secondary KV cache pool, designers can flexibly choose either block-level or session-level eviction, since the latter drastically reduces metadata overhead without incurring a performance penalty.

\section{Discussion}

\subsection{Other Attention Designs}
\label{sec:discussion:attention}

While blocks are uniformly sized, the underlying attention mechanism dictates a highly variable recomputation cost.
Our three cost models ablate this dynamic, spanning from a uniform-cost baseline (\S\ref{sec:algorithms:evaluation}) to the \model cost (\S\ref{sec:designs:compute}) and an idealized linear model (Appendix~\ref{sec:appendix:linear}).
Alternative attention designs dictate where a given deployment falls within this spectrum.
Sliding-window attention~\cite{jiang2023mistral} caps cost growth with prefix depth and moves the
objective closer to constant cost, while in hybrid attention models~\cite{lieber2024jamba}, the benefit
scales with the share of full-attention layers.
GQA~\cite{ainslie2023gqa}, MQA~\cite{shazeer2019mqa},
MLA~\cite{deepseekai2024deepseekv2}, cross-layer KV
sharing~\cite{brandon2024cla}, and KV quantization~\cite{liu2024kivi} can also
change recomputation cost. These designs require recalibrating the cost model, while
the reuse patterns in \S\ref{sec:analysis} still apply.

\subsection{Interaction with Load Balancing}
\label{sec:discussion:lb}

While routing load balancers often use cache affinity to raise local reuse, they can create uneven cache pressure across replicas due to the heavy-tailed session footprints we identified (\S\ref{sec:analysis:footprint})~\cite{wang2026smetric,hayes2026accelerate}. Regardless of how a router distributes these requests across the cluster, our findings and policy designs remain fully applicable. Once a request lands on a specific replica or shared memory pool, the underlying access patterns---session-bounded lifetimes (\S\ref{sec:analysis:lifetime}), concentrated recency (\S\ref{sec:analysis:reuse}), and varying compute costs (\S\ref{sec:analysis:miss-penalty})---still strictly dictate the cache dynamics. Consequently, our recency-based eviction insights, quick demotion, and compute-aware optimizations hold true as the foundational eviction layer beneath any load balancing strategy.

\subsection{SSD Offloading}
\label{sec:discussion:ssd}

Given the extremely short, session-bounded lifetimes of prefix-cache blocks (\S\ref{sec:analysis:lifetime}), offloading evicted blocks from HBM to an SSD tier~\cite{liu2025lmcache} is generally unnecessary for most workloads, as the working set churns away before the SSD capacity provides any benefit. However, in specific cases where workloads exhibit long-tail reuse or extreme capacity demands, SSD offloading may still be deployed. In these scenarios, a practical SSD tier must employ strict selective admission to preserve limited flash endurance~\cite{schroeder2016flash} while capturing the narrow window of actual reuse~\cite{eisenman2019flashield,mcallister2021kangaroo}.

\section{Related Work}

\pgheading{Prefix caching and storage}
vLLM~\cite{kwon2023efficient} and SGLang~\cite{zheng2024sglang} enable prefix reuse
via paged and radix-tree KV caches. Mooncake,
Strata, and PTStore~\cite{qin2025mooncake,xie2025strata,madhyastha2026ptstore}
extend KV storage across GPU memory, host memory, SSDs, and remote nodes.
CachedAttention and IMPRESS~\cite{gao2024cachedattention,chen2025impress}
manage retention across these tiers. We study the replacement policy within
such mechanisms and how it should change with the memory tier.

\pgheading{Prefix-cache eviction}
Recent policies predict session continuation, partition workloads, or rank
blocks using recency, frequency, and semantic features
\cite{yang2025learned,ouyang2026unicache,li2025hotprefix,fang2026saecache,wang2025kvcache,shi2026asymcache}.
Marconi~\cite{pan2025marconi} models compute saved per byte for hybrid
attention models. We instead compare classical, learned, and
prefix-specific policies over two production traces and a wide capacity
range, then explain their behavior through session-bounded reuse, frequency
(accesses per object), non-uniform miss cost, and session footprint. These
properties motivate our compute-aware and tier-dependent designs.

\pgheading{Adjacent cache decisions}
CacheWise~\cite{tiwari2026cachewise} and KVFlow~\cite{pan2025kvflow} use tool or
workflow metadata unavailable in the access stream to predict reuse.
Continuum~\cite{li2025continuum}, InferCept~\cite{abhyankar2024infercept},
and EvicPress~\cite{feng2025evicpress}
decide whether to preserve or move one request's context, while cache-aware
routers~\cite{wang2026smetric,hayes2026accelerate} decide where a request
runs. These signals and decisions complement eviction, which determines what
remains available at each replica.

\pgheading{Workload characterization}
Production studies characterize prompt composition, sessions, and observed
cache reuse~\cite{liu2026agentic,zhu2026tracelab,wang2025kvcache,aubakirova2026state}.
However, existing public traces lack the combination of agentic traffic, accurate multi-day timestamps, and fine-grained block hashes necessary for robust policy simulation.
By introducing traces with these properties, our controlled trace replay enables comparing eviction policies and their offline optimum on the same request stream.

\pgheading{Approximate KV-cache management}
H2O~\cite{zhang2023h2o}, Scissorhands~\cite{liu2023scissorhands},
StreamingLLM~\cite{xiao2024streaming}, and SnapKV~\cite{li2024snapkv} remove
KV entries within one active context; Quest~\cite{tang2024quest} and
InfiniGen~\cite{lee2024infinigen} skip or prefetch them. These
methods may trade output quality for per-request memory, whereas our
cross-request blocks are exactly reconstructible and a miss costs prefill
compute. Our evaluated policies draw from classical recency, frequency,
admission, adaptive, analytic, and learned replacement
\cite{yang2023fifo,megiddo2003arc,jiang2002lirs,beckmann2018lhd,vietri2018lecar,song2020learning,zhou20253lcache},
with Belady's algorithm as the offline optimum~\cite{belady1966study}.

\section{Conclusion}
\label{section:conclusion}

As LLM serving scales, prefix caching has become essential to bypass expensive prefill recomputations, yet we find that advanced eviction algorithms built for traditional storage yield surprisingly little improvement over simple LRU. Our systematic trace analysis uncovers the root cause: prefix reuse follows the steady pacing of active sessions, making recency very crucial and rendering frequency-based signals ineffective. We also identify unique challenges in prefix caching: highly non-uniform miss costs that scale with depth and heavy-tailed session footprints. To optimize beyond simple hit ratios, systems must retain a recency foundation but augment it conditionally: deploying quick demotion to filter one-hit wonders, introducing compute-aware eviction to prioritize expensive blocks, and adapting eviction granularity to the capacity tier. Together, these targeted optimizations provide a concrete foundation for scalable prefix caching.

\bibliographystyle{plain}
\bibliography{reference}

\appendix
\section{Capacity Sweeps}
\label{sec:appendix:more-traces}

\begin{figure*}[t]
  \centering
  \begin{subfigure}{0.48\linewidth}
    \centering
    \includegraphics[width=\linewidth]{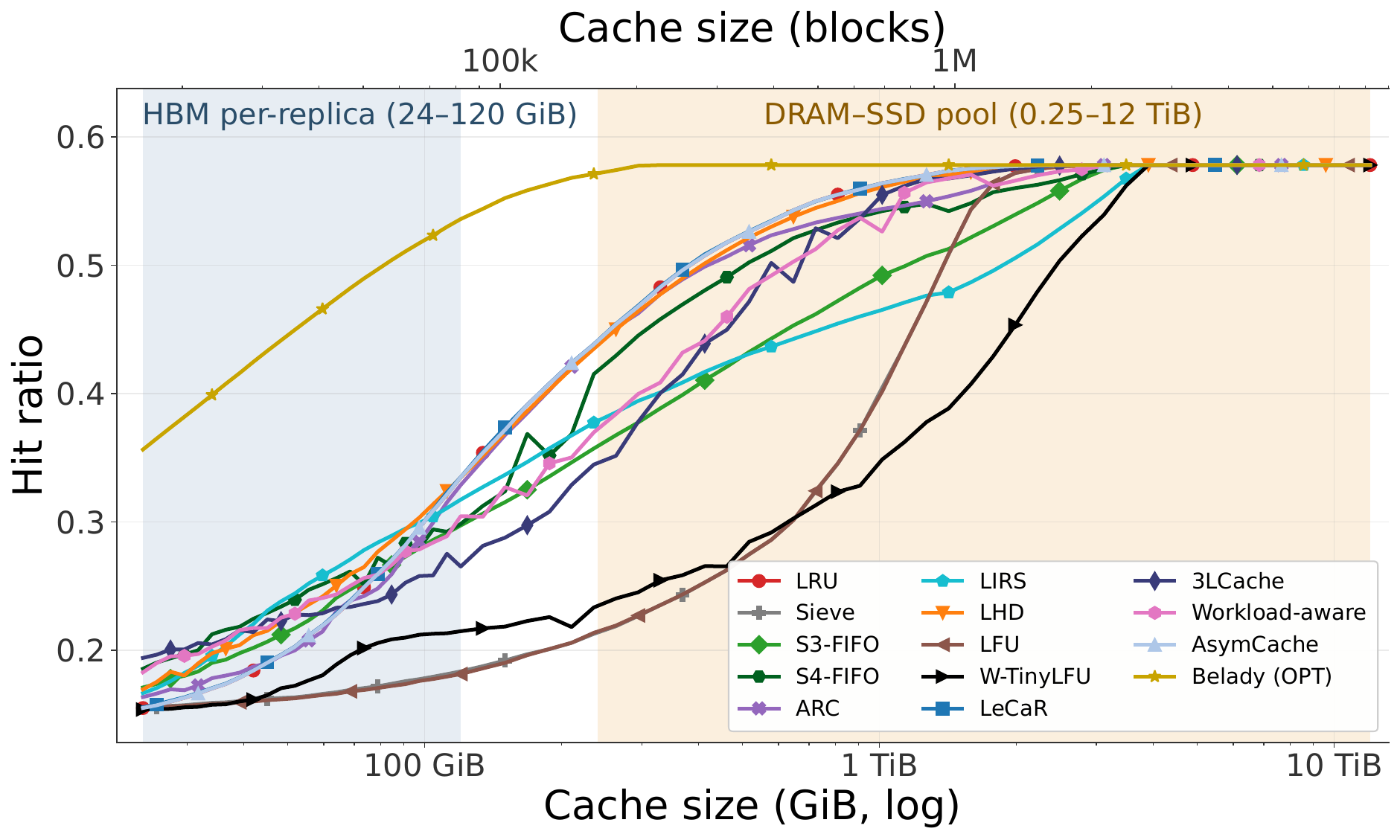}
    \caption{Qwen To-C.}
    \label{fig:sweep-traceA}
  \end{subfigure}
  \hfill
  \begin{subfigure}{0.48\linewidth}
    \centering
    \includegraphics[width=\linewidth]{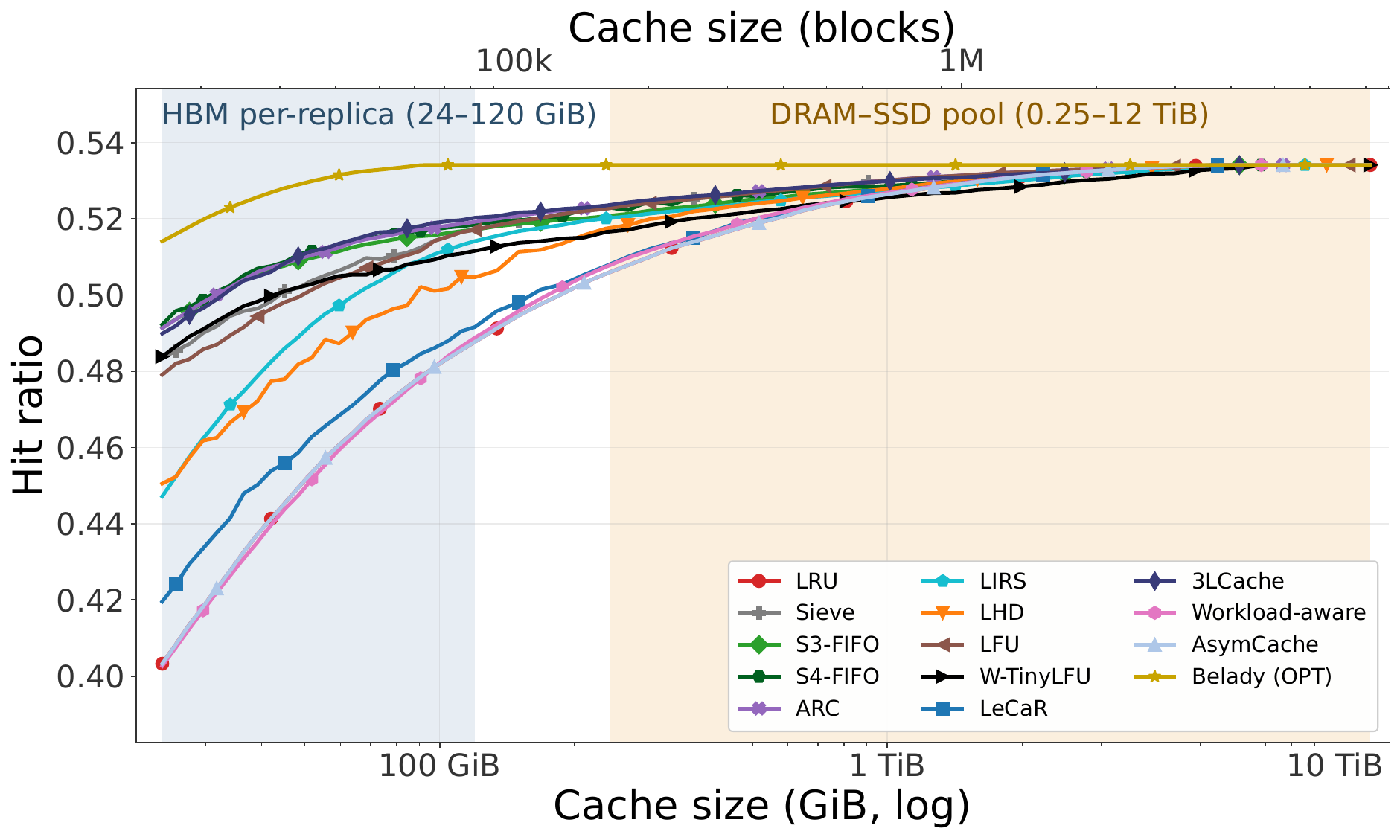}
    \caption{Qwen To-B.}
    \label{fig:sweep-traceB}
  \end{subfigure}

  \vspace{0.6em}
  \begin{subfigure}{0.48\linewidth}
    \centering
    \includegraphics[width=\linewidth]{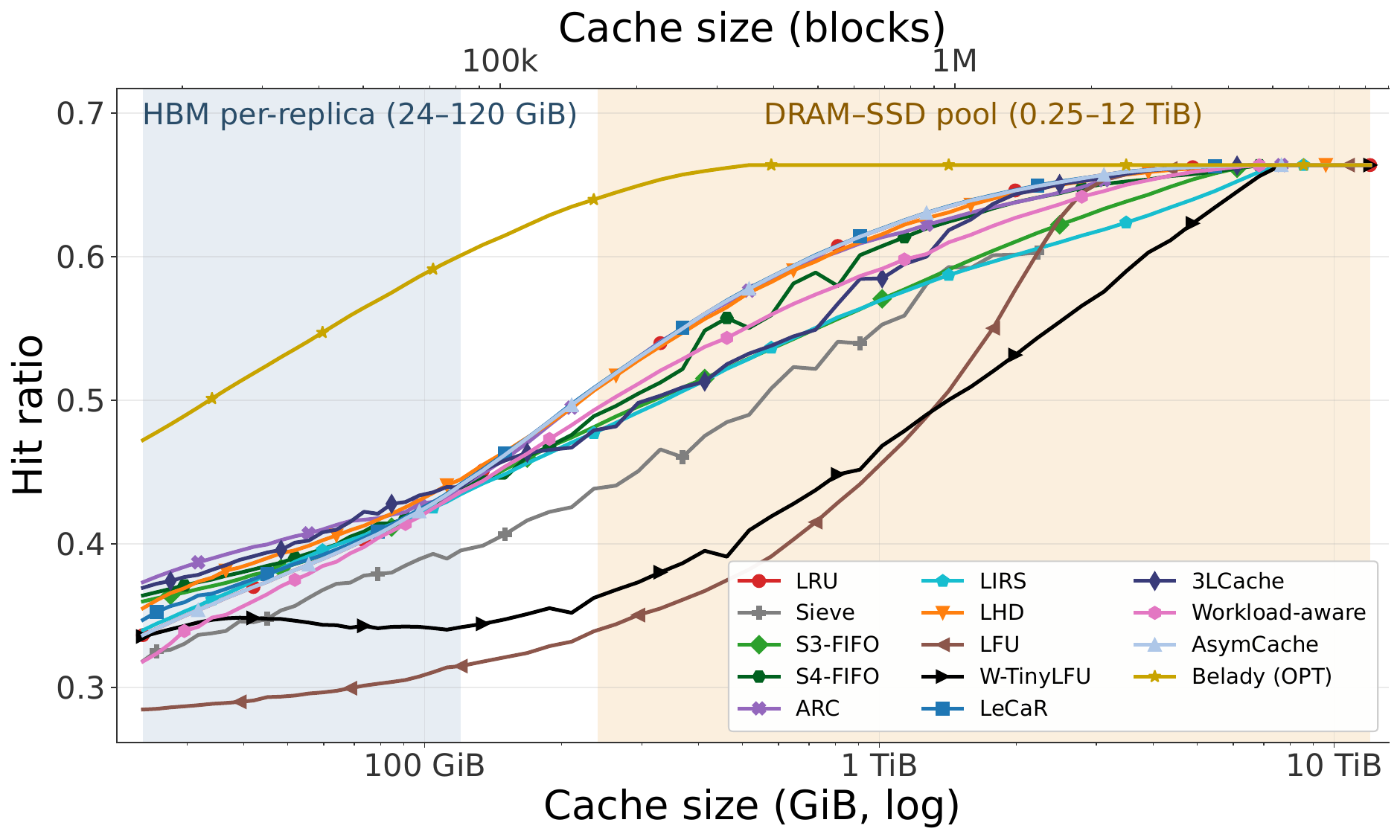}
    \caption{Qwen Coder.}
    \label{fig:sweep-coder}
  \end{subfigure}
  \hfill
  \begin{subfigure}{0.48\linewidth}
    \centering
    \includegraphics[width=\linewidth]{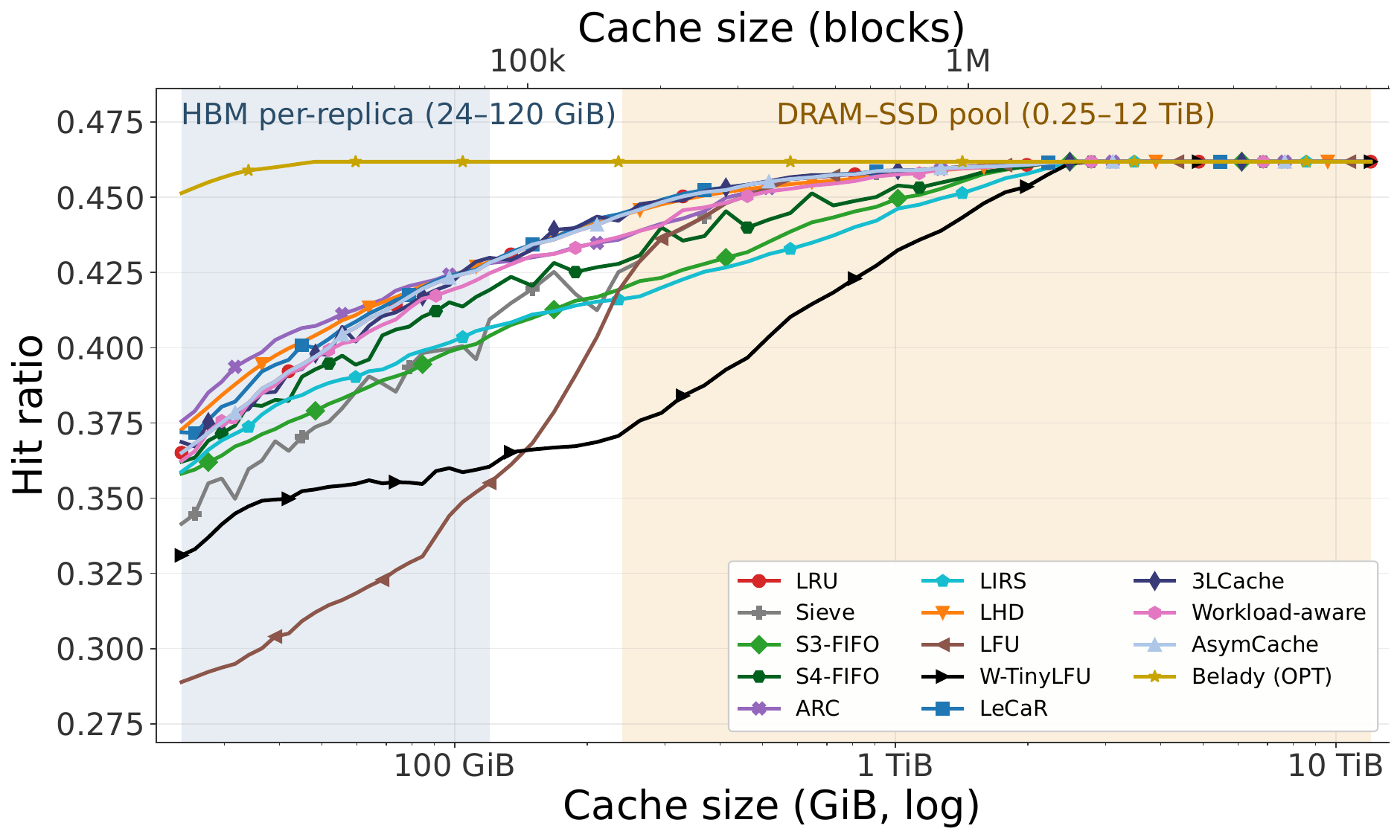}
    \caption{Qwen Thinking.}
    \label{fig:sweep-thinking}
  \end{subfigure}
  \caption{Block hit ratio against cache size on the four Qwen Bailian
  traces, over the two cache-size regimes of
  \S\ref{sec:algorithms:regimes}. Figure~\ref{fig:bars8-qwen} reports the
  corresponding comparison at four capacities.}
  \label{fig:sweep-qwen}
\end{figure*}

\begin{figure*}[t]
  \centering
  \begin{subfigure}{0.48\linewidth}
    \centering
    \includegraphics[width=\linewidth]{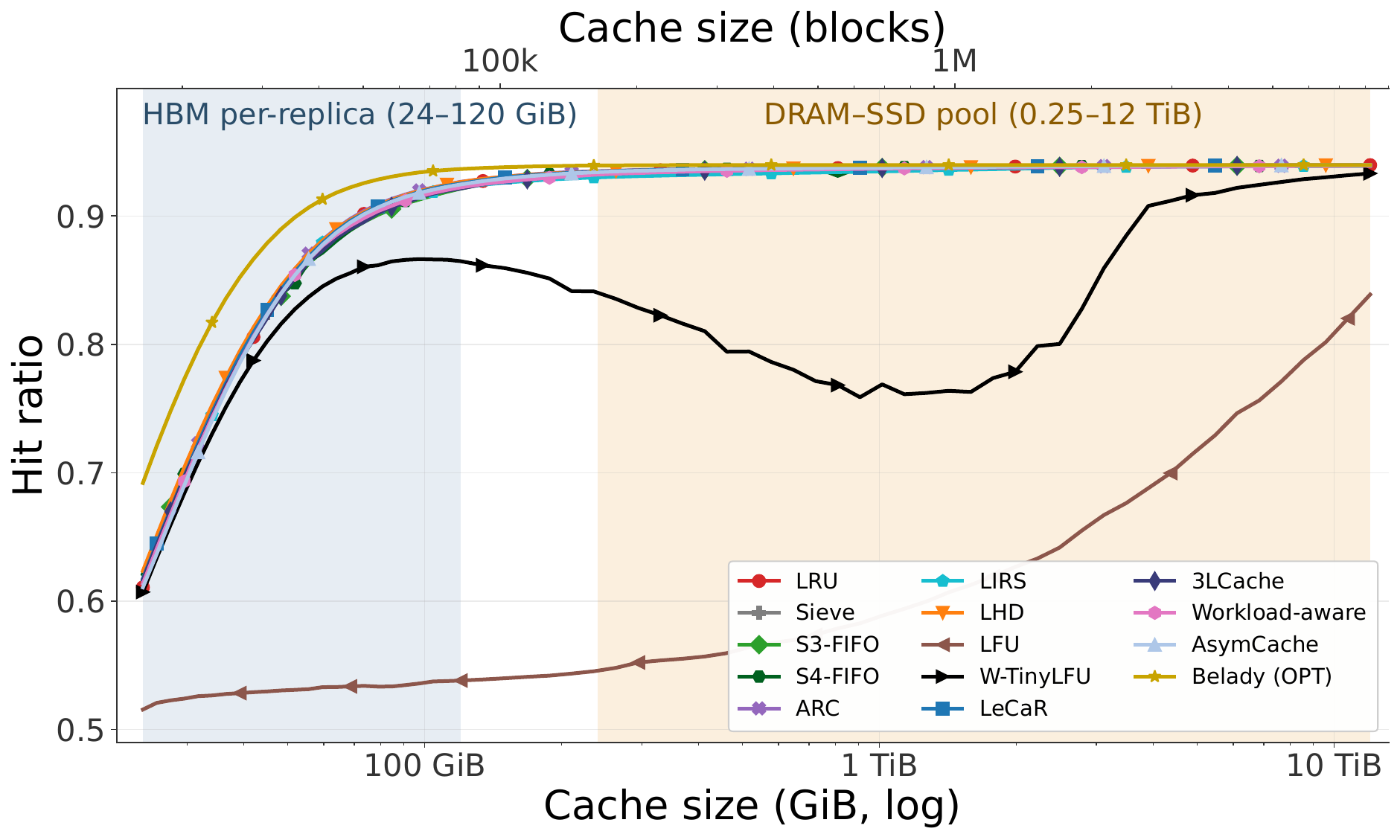}
    \caption{\tracea.}
    \label{fig:sweep-fi}
  \end{subfigure}
  \hfill
  \begin{subfigure}{0.48\linewidth}
    \centering
    \includegraphics[width=\linewidth]{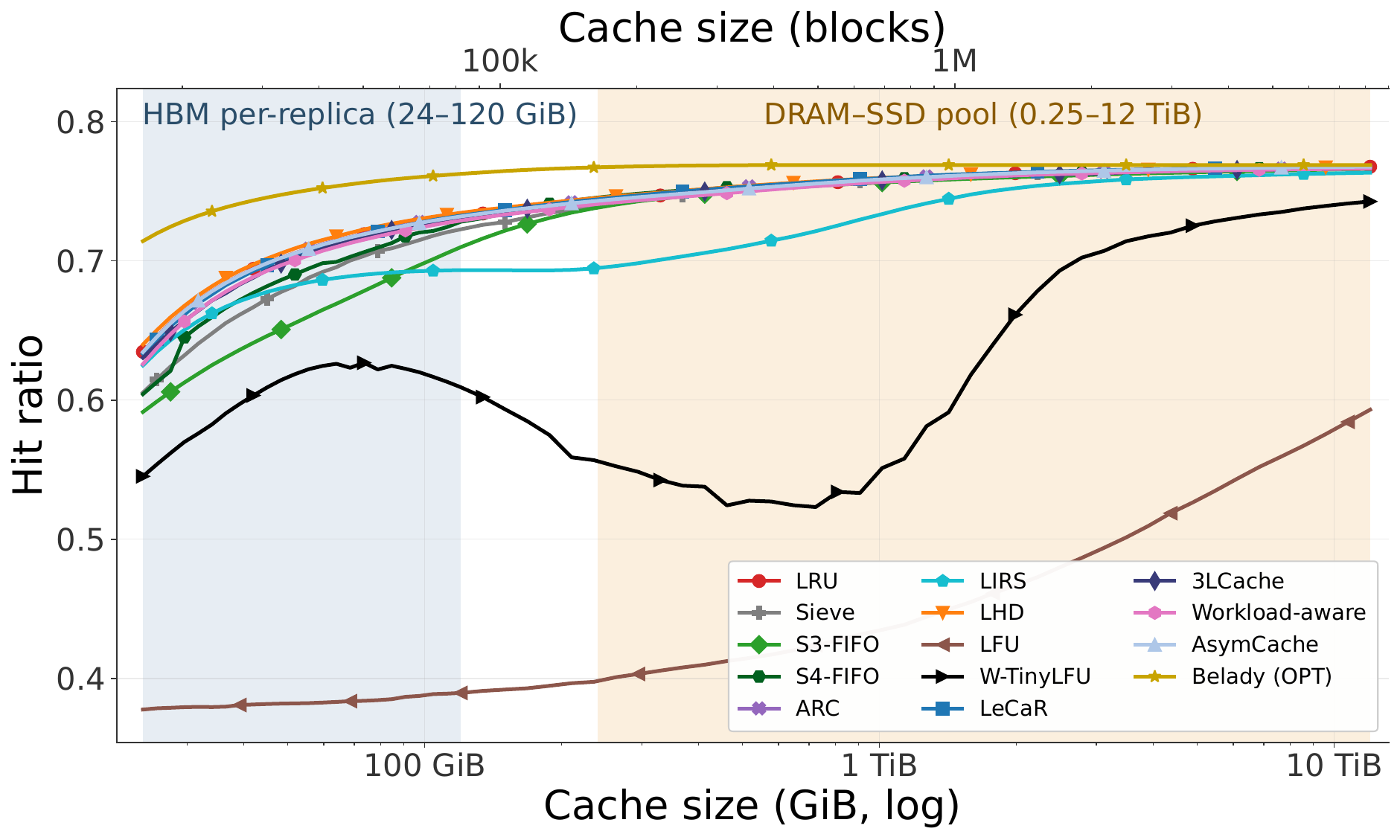}
    \caption{\traceb.}
    \label{fig:sweep-chutes}
  \end{subfigure}
  \caption{Block hit ratio against cache size on the two production traces of
  \S\ref{sec:algorithms:evaluation}. Figure~\ref{fig:bars-traces} reports the
  corresponding comparison at four capacities.}
  \label{fig:sweep-prod}
\end{figure*}

Sections~\ref{sec:analysis:more} and~\ref{sec:algorithms:evaluation} compare
policies at four cache capacities on six traces. Figures~\ref{fig:sweep-qwen}
and~\ref{fig:sweep-prod} extend this comparison to 65 capacities across both
cache-size regimes. Both sweeps include every policy in
Figure~\ref{fig:bars-traces} except LRB. We omit LRB because it is prohibitively slow
to sweep and never outperforms LeCaR at the four sampled capacities on either
production trace.

\begin{figure}[t]
  \centering
  \includegraphics[width=\linewidth]{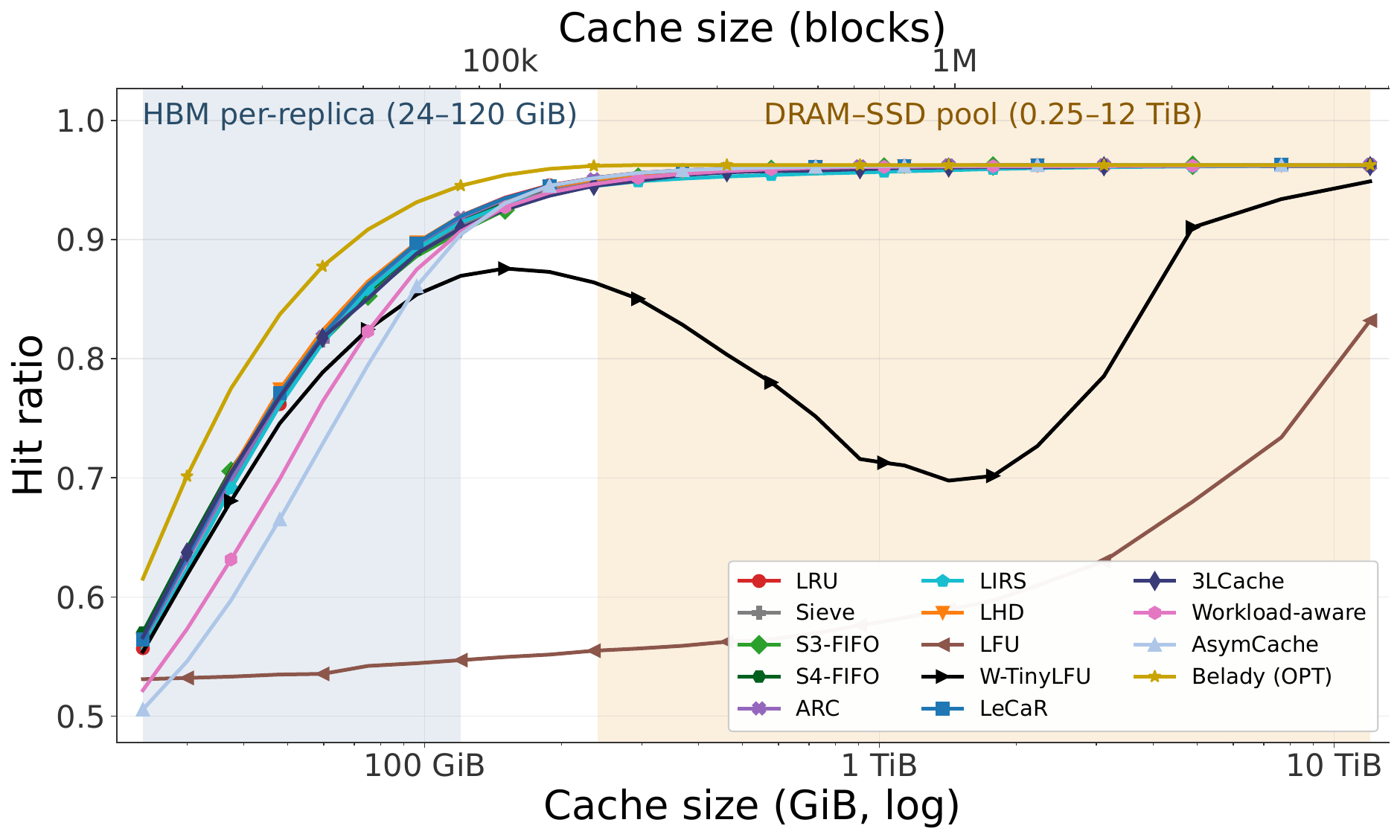}
  \vspace{-0.5em}
  \caption{Block hit ratio against cache size on AgentX. Sampled at 26
  capacities over the same range as
  Figures~\ref{fig:sweep-qwen} and~\ref{fig:sweep-prod};
  Figure~\ref{fig:bars8-qwen} reports the corresponding comparison at four
  capacities.}
  \label{fig:sweep-agentx}
\end{figure}

\begin{table}[t]
  \centering
  \caption{Sizes of the four Qwen Bailian traces and the two production
  traces.}
  \label{tab:secondary-traces}
  \small
  \begin{tabular}{@{}l rr@{}}
    \toprule
    Trace & Requests & Block accesses \\
    \midrule
    Qwen To-C      &  43.1\,K &   6.3\,M \\
    Qwen To-B      & 172.8\,K &  10.0\,M \\
    Qwen Coder     &  43.0\,K &  15.5\,M \\
    Qwen Thinking  &  10.8\,K &   3.2\,M \\
    \midrule
    \tracea        & 327.5\,K & 656.2\,M \\
    \traceb        & 515.8\,K & 603.6\,M \\
    \bottomrule
  \end{tabular}
\end{table}

The denser sweeps expose two behaviors hidden by the four sampled capacities.
First, the gains from quick demotion can reverse as capacity grows. On Qwen
To-C, for example, LIRS moves from 4.2 points above LRU to 10.1
points below it at 724\,GB before converging again. Second, W-TinyLFU is
nonmonotonic in cache size. On \tracea, its hit ratio rises to 0.866 at 97\,GB,
falls to 0.759 at 906\,GB, and recovers to 0.933 at 12\,TB. Thus, comparisons
at a few capacities can hide both policy crossovers and nonmonotonic behavior.

\section{Sensitivity Analysis of Concurrency}
\label{sec:appendix:concsens}

\begin{figure}[t]
  \centering
  \includegraphics[width=\linewidth]{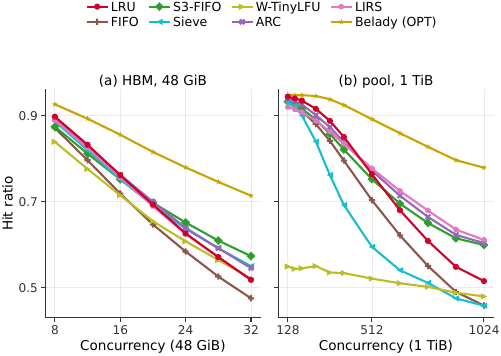}
  \caption{Hit ratio against session concurrency at a fixed cache size, on the
  synthetic multi-turn trace, in the per-replica HBM regime (a) and the
  DRAM/SSD pool regime (b); the two panels share a $y$ axis. Every point
  replays the same set of requests.}
  \label{fig:concsens}
\end{figure}

\tracea has relatively low session concurrency, averaging about 12 concurrent
sessions, and every policy except W-TinyLFU remains within roughly one point
of LRU across the capacity sweep (Figure~\ref{fig:sweep-fi}). To isolate the
effect of concurrency, we fix the cache at 48\,GiB in HBM and 1\,TiB in the
pool, then synthetically vary session concurrency
(Figure~\ref{fig:concsens}). We extract the multi-turn sessions from \tracea,
truncate them where concurrency begins to collapse in the 1024-concurrency
replay, and rescale their arrival times at each target concurrency. Every
point therefore replays the same 73{,}637 requests, with concurrency as the
only varied factor. Because truncation retains the earlier portion of each
session, the absolute hit ratios should not be compared with
\S\ref{sec:algorithms:regimes}; the trend across concurrency is the relevant
result.

Hit ratio declines for every policy as concurrency increases, including the
offline optimum. At 48\,GiB, Belady falls from 0.926 at concurrency 8 to 0.713
at concurrency 32, while LRU falls from 0.898 to 0.517.

Policy differences also widen because concurrency increases the stack distance
between reuses. In the HBM regime, the fraction of reuses reachable within
48\,GiB falls from 92.5\% at concurrency 8 to 50.1\% at concurrency 32. Blocks
that will be reused later in a multi-turn session thus behave as one-hit blocks
within the cache horizon, making quick demotion more effective
(\S\ref{sec:designs:qd}). At concurrency 32, S3-FIFO reaches 0.573, compared
with LRU's 0.517. The pool regime shows the same effect: at concurrency 1024,
only 54.1\% of reuses are reachable within 1\,TiB, and S3-FIFO, ARC, and LIRS
overtake the other online policies. W-TinyLFU's flatter curve in this regime
is a floor effect. Its frequency filter already discards useful blocks when
capacity is abundant, leaving less performance to lose as concurrency rises.

\section{Compute Savings under the Linear Cost Model}
\label{sec:appendix:linear}

\begin{figure}[t]
  \centering
  \includegraphics[width=\linewidth]{figures/compute_bars_legend.pdf}\\[2pt]
  \includegraphics[width=\linewidth]{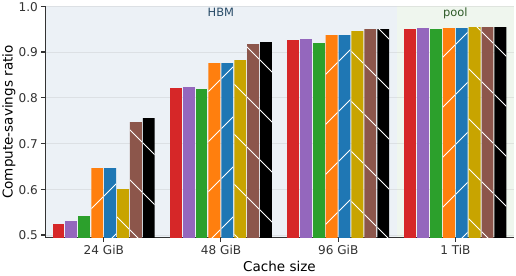}
  \vspace{-0.5em}
  \caption{Compute-savings ratio on the prefix-cache trace under an idealized
  linear cost, where a block's recompute cost is proportional to its position in
  the prompt. Same policies, colors and capacities as
  Figure~\ref{fig:compute-aware}, which uses the measured cost model.}
  \label{fig:compute-aware-linear}
\end{figure}

\S\ref{sec:designs:compute} reports compute savings under the measured
\model\ cost model. Figure~\ref{fig:compute-aware-linear} repeats that
measurement under a linear cost, the extreme case in which the attention
matrix multiplication dominates and a block's recompute cost is proportional to
its depth in the prompt. The two models rank the policies identically, and
every claim in \S\ref{sec:designs:compute} holds here with a slightly wider
margin.

Compute-awareness pays more under the linear model, as expected: it is the
model with the larger spread in per-block cost, because the measured profile
adds a large position-independent term that shallow and deep blocks both pay.
At 24\,GiB, RandomCompute saves 0.646 against LRU's 0.524, a margin of 12.2
points where the measured model gives 10.4, and it clears the hit-optimal
Belady oracle by 4.7 points rather than 2.9. The gap between the two oracles
widens the same way: BeladyCompute reaches 0.745 against Belady's 0.599, or
14.6 points, against 12.6 under the measured model.

BeladyCompute remains a close approximation of the offline optimum: it is 0.95
points below the ILP solution at 24\,GiB, 0.40 at 48\,GiB and 0.08 at
96\,GiB, matching the measured model to within 0.1 points at every cache size.
Partial-node eviction stays close to free under both: at 24\,GiB it gives up
0.17 points of compute savings under the measured model and 0.03 under the
linear one, so the fragmentation fix does not become more expensive when deep
blocks are weighted more heavily. Past the per-replica regime the choice of cost model
stops mattering: at 1\,TiB every policy in the figure, both oracles included,
lies within 0.2 points of the others.

\section{Recompute Fragmentation}
\label{sec:appendix:holes}

\begin{figure}[t]
  \centering
  \includegraphics{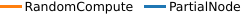}\\[-0.5em]
  \begin{subfigure}[t]{0.48\linewidth}
    \vspace{0pt}
    \centering
    \includegraphics[width=\linewidth]{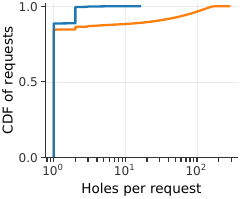}
    \caption{Holes per request.}
    \label{fig:holes-per-request}
  \end{subfigure}
  \hfill
  \begin{subfigure}[t]{0.48\linewidth}
    \vspace{0pt}
    \centering
    \includegraphics[width=\linewidth]{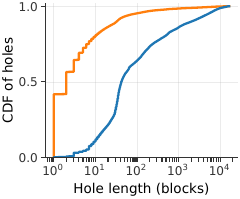}
    \caption{Hole length.}
    \label{fig:holes-length-cdf}
  \end{subfigure}
  \caption{Effect of partial-node eviction on recompute fragmentation at
  24\,GiB over the 327{,}548 requests of \tracea. Partial-node eviction
  leaves fewer holes per request (a), and those holes are longer (b).}
  \label{fig:holes}
\end{figure}

Figure~\ref{fig:holes} gives the distributions behind the averages reported in
\S\ref{sec:designs:compute}. Both policies have a median of one hole per
request, but RandomCompute's tail reaches 143 holes at the 99th percentile and
281 at the maximum, compared with 2 and 16 under partial-node eviction. The
total recompute volume is nearly unchanged, at 258.0\,M and 257.9\,M blocks,
respectively. Partial-node eviction instead packs this work into longer runs:
mean hole length rises from 85.5 to 742.6 blocks, while the fraction of holes
no longer than two blocks falls from 56.6\% to 0.8\%.

\end{document}